# Forest tree species classification and entropy-derived uncertainty mapping using extreme gradient boosting and Sentinel-1/2 satellite data

Abdulhakim M. Abdi * and Fan Wang
*Centre for Environmental and Climate Science, Lund University, Sweden*

* hakim.abdi@cec.lu.se

## Abstract

Timely, detailed information on forest composition is essential for effective management, biodiversity protection, and understanding ecosystem dynamics. As forest assessments move toward finer spatial detail, approaches that characterize tree species and the confidence of those classifications have become particularly useful. This study maps seven dominant tree species in Swedish forests and produces spatially explicit, pixel-level estimates of classification uncertainty. The mapping framework integrates multitemporal Sentinel-1 radar and Sentinel-2 optical observations with field data from the Swedish National Forest Inventory and auxiliary predictors describing geomorphometry and canopy height. We trained a Bayesian-optimised extreme gradient boosting model on spatiotemporal metrics derived from these datasets. We quantified classification confidence through entropy computed from the class-probability outputs. To limit the effects of spatial autocorrelation between optimisation and validation data, we applied a spatial block partitioning approach, ensuring a more realistic assessment of the model's generalization capacity. The final model achieved an overall accuracy of 85% (F1 score = 0.82), and the county-level species coverage derived from the classification aligned closely with published figures from the Swedish Forest Agency (Spearman's $\rho = 0.94$, $P < 0.05$). Variable importance analysis showed that Sentinel-2 spectral information, particularly shortwave-infrared and red-edge bands captured during spring and summer, contributed most to species discrimination, while Sentinel-1 backscatter provided complementary structural cues. The integration of satellite time series, forest inventory data, and an explicit measure of prediction uncertainty yields a reproducible framework for large-area forest species mapping. The resulting products provide detailed, spatially continuous information on species composition along with an accompanying confidence surface, offering practical value for ecological assessments, regional planning, and emerging legislative and environmental goals.

## Introduction

Forests play a central role in global carbon sequestration (Pan et al. 2024), sustain biodiversity (Peh, Corlett, and Bergeron 2025), and support human well-being (Dlamini 2019; Creed et al. 2016). These ecosystem services depend on species-specific traits (Brockerhoff et al. 2017), such as growth rates, shade tolerance, and disturbance responses, making knowledge of tree species composition and distribution essential for sustainable management. Tree species differ in how they influence ecosystem processes, and greater tree species diversity has been shown to enhance productivity, stability, and biodiversity outcomes (Morin et al. 2018; Morin et al. 2011; Liang et al. 2016). Individual tree species also shape environmental conditions, for example by modifying the microclimate (Zhang et al. 2022) or affecting soil carbon storage (Liu et al. 2018). Therefore, detailed knowledge of tree species distributions is essential for informing management decisions and supporting the ecological and societal benefits that forests provide.

Sweden has the largest forest area in the European Union, covering approximately 280,000 $km^2$, or about 70% of its land area (Roberge et al. 2020). Forests are an integral part of Sweden's economy, providing resources such as timber and non-timber products that make it the world's third-largest exporter of coniferous sawn wood and second-largest exporter of paper and pulp (Roberge et al. 2020). Ninety-two percent of the growing stock in Swedish forests comprises of just three dominant species – Norway spruce (*Picea abies*, 40.3%), Scots pine (*Pinus sylvestris*, 39.3%), and Birch (*Betula spp.*, 12.4%). Although Sweden covers a wide biogeography, ranging from boreal to nemoral/continental (Preislerová et al. 2024), the dominance of these three species is high in northern parts of the country (97% of the growing stock) and low in the southern parts (86%). Thus, the highest levels of tree diversity occur in southern Sweden (Götaland, Figure 1) where most of the deciduous forests comprise Oak (*Quercus spp.*), Beech (*Fagus sylvatica*), and other nemoral broadleaved trees (Roberge et al. 2020).

The Swedish National Forest Inventory (NFI) serves as a primary source for deriving statistics on forest resources at both national and subregional scales (Fridman and Westerlund 2016). The inventory area is divided into geographic strata to ensure adequate representation of different forest types, land uses, and regions. Although the NFI spans the entire country, its plot-based sampling design and multi-year survey cycle limit its value for fine-scale spatiotemporal analysis. In addition, the exact plot locations are withheld to reduce the risk of disturbance and maintain the reliability of the data (Schadauer et al. 2024). These limitations necessitate the use of complementary methods, such as Earth observation (EO) and machine learning (ML), that can be deployed in a timely manner to efficiently map Swedish forests.

The past decade has seen a broad push within Europe toward more harmonised and data-rich approaches to forest monitoring, driven by concerns about accelerating disturbance regimes, declining biodiversity, and rising management pressures (Atzberger et al. 2020; European Commission 2021; Raši 2020; Patacca et al. 2023). These discussions highlight a persistent gap – that field-based monitoring systems alone struggle to capture rapid, spatially heterogeneous changes occurring across large, forested landscapes. As a result, there is growing emphasis on digital forest information (Nitoslawski et al. 2021), timely disturbance reporting (Senf and Seidl 2021; Viana-Soto and Senf 2025), and the integration of Earth observation into national monitoring frameworks (Cesaretti et al. 2025). These developments highlight the value of scalable remote sensing methods that complement conventional forest inventories. Given Sweden's extensive forest cover and reliance on sample-based inventories, satellite-derived tree species mapping provides a practical way to supply spatially explicit forest information that can help meet emerging legislative and policy demands.

EO technology provides valuable tools to address the challenges of large-scale forest species mapping, particularly when integrated with field-based NFIs. Canada provides a good example of such integration; for example Hermosilla, Bastyr, et al. (2022) mapped the distributions of 37 tree species across Canada's forests using a harmonised EO framework validated with NFI data. Similar national initiatives have been implemented in Europe. In Norway, Breidenbach et al. (2021) combined Sentinel-2 data and NFI measurements to produce nationwide maps of dominant tree species and forest area. In Germany, three nationwide forest species maps have been developed using Sentinel-1/2 time series – two with NFI data (Welle et al. 2022; Blickensdörfer et al. 2024) and one without (Wegler et al. 2025). And in Poland, Grabska-Szwagrzyk et al. (2024) produced the country's first forest tree species map using Sentinel-2 time series and data from the Polish NFI. Collectively, these national efforts demonstrate the growing trend toward combining EO and NFI data to produce spatially explicit, policy-relevant forest information.

In recent years, the Copernicus Sentinel missions have become key resources for large-scale forest monitoring. Sentinel-1, carrying a C-band synthetic aperture radar (SAR), and Sentinel-2, carrying an optical Multispectral Instrument (MSI), together provide complementary information on forest structure and composition. The Sentinel-2 constellation currently includes three operational satellites – Sentinel-2A (launched in June 2015), Sentinel-2B (March 2017), and Sentinel-2C (September 2024), which collectively provide a nominal 5-day revisit cycle at the equator and higher temporal frequency (2 to 3 days) at northern latitudes. The spatial resolution provided by MSI ranges from 10 to 60 m. The Sentinel-1 constellation, comprising Sentinel-1A (launched in April 2014), Sentinel-1C (December 2024), and Sentinel-1D (November 2025), provides SAR backscatter data at 5 – 20 m resolution with a 6-day revisit interval. Sentinel-1B, launched in 2016, ceased operations in August 2022 following a power supply failure. Together, these satellite instruments have high thematic accuracy for tree cover mapping (Ottosen et al. 2020; Lechner et al. 2022). They have also proven effective in mapping forest type and tree species across different European ecological zones, for example nemoral (Grabska et al. 2019), hemiboreal(Udali, Lingua, and Persson 2021), forest-steppic (Vorovencii et al. 2023), and Mediterranean (Puletti, Chianucci, and Castaldi 2018).

The combined use of Sentinel-1 and Sentinel-2 has been strengthened by the development of cloud computing infrastructures, such as Microsoft Planetary Computer (Joppa 2019) and Google Earth Engine (GEE) (Gorelick et al. 2017), that enable the generation of multitemporal composites. Spectral-temporal metrics (STMs) derived from such composites summarize the statistical properties of each pixel (e.g., mean, standard deviation, percentiles) over specific time periods, such as seasonal or annual windows. This approach condenses large image stacks into a single, information-rich composite, capturing key phenological and structural variations while mitigating issues related to cloud cover and atmospheric variability. STMs have proven effective for large-scale tree species mapping (Grabska-Szwagrzyk et al. 2024; Blickensdörfer et al. 2024) and form an essential foundation for contemporary ML applications in forest classification.

In this study, we mapped and quantified the pixel-level uncertainty of seven dominant tree species (Table 1) that together represent 97% of the growing stock in Sweden's forests (Riksskogstaxeringen 2024), at a spatial resolution of 10 m. The objectives were to (i) assess the predictive capacity of Sentinel-1 and Sentinel-2 data, including spectral bands, polarization channels, vegetation indices, and STMs, for tree species classification, and (ii) evaluate the influence of geomorphometric variables (slope, aspect, elevation, topographic wetness) and canopy height on species distributions. Field data from the NFI were used for model optimisation and validation. We used extreme gradient boosting (Chen and Guestrin 2016) with Bayesian optimisation to generate species distribution maps and derived pixel-level uncertainty using Shannon's entropy (Wang 2008). The final maps were compared with official forest statistics (Riksskogstaxeringen 2024) to assess spatial consistency. Overall, the study aims to deliver an accurate and interpretable framework for tree species mapping and uncertainty quantification, providing a step toward operational forest monitoring.

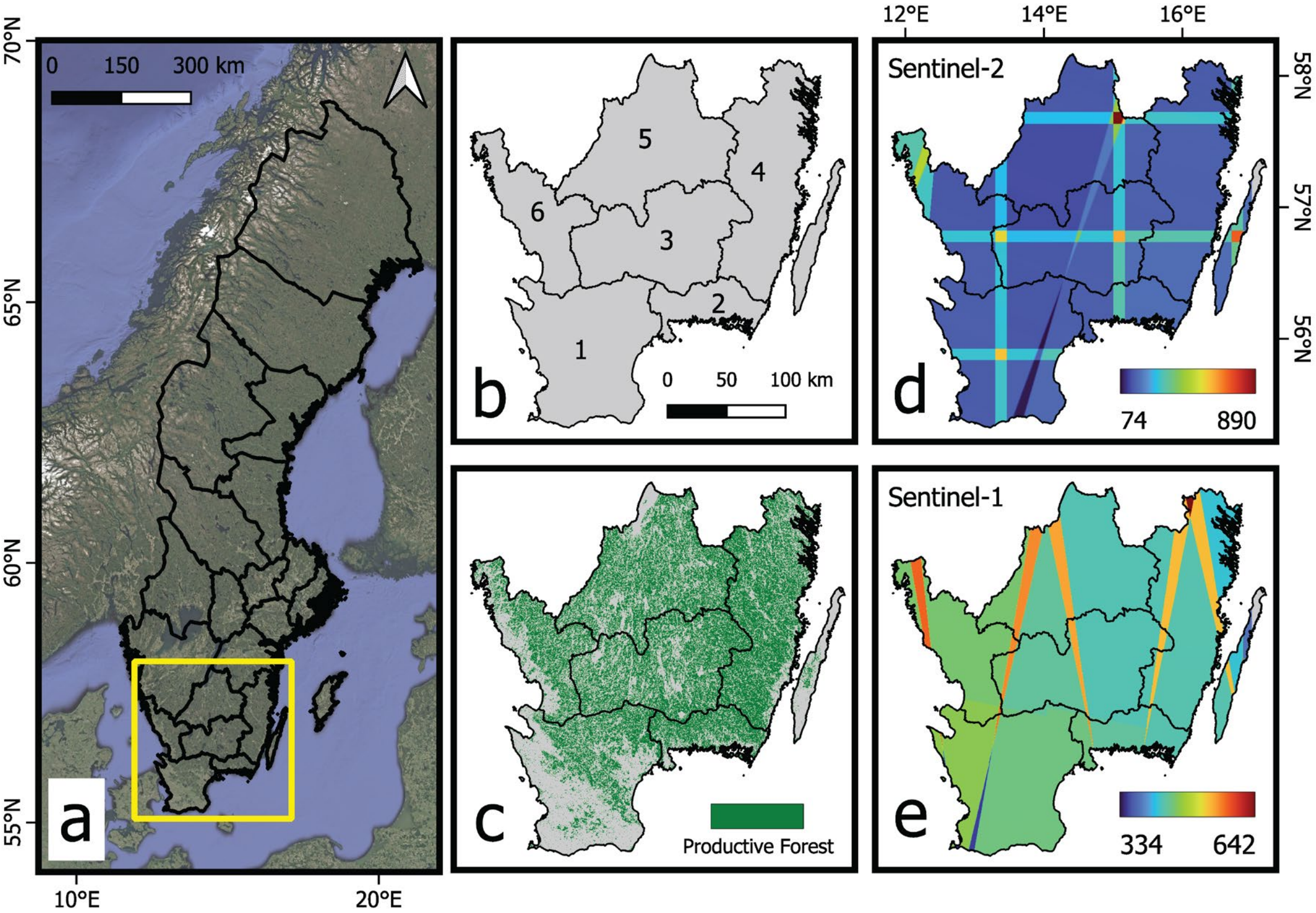

**Figure 1.** The study area in southern Sweden **(a)** is composed of six counties (*län*) **(b)**, Skåne (b1), Blekinge (b2), Kronoberg (b3), Kalmar (b4), Jönköping (b5), and Halland (b6), and most of it is classified as productive forest according to the Swedish National Land Cover Database **(c)**. The final two panels display the coverage of satellite observations between 2021 and 2023 for Sentinel-2 MSI with less than 70% cloud cover **(d)** and Sentinel-1 SAR in both orbital directions **(e)**.

## Study area

The study area encompasses the productive forests of southern Götaland in south of Sweden (Figure 1a) and includes six administrative counties (Figure 1b) covering 52,120 km². Under Sweden's Forestry Act (Riksdag 1979), productive forest refers to forested land capable of producing, on average, at least one cubic meter of timber per hectare annually. Accordingly, the analysis was spatially limited to areas classified as productive forest in the National Land Cover Map, which represent 57% of the region (Figure 1c). Ecologically, Götaland spans hemiboreal and nemoral zones, forming a transition from the conifer-dominated boreal systems of northern Sweden to the broadleaf-rich temperate forests characteristic of continental Europe (Roberge et al. 2020). In northern parts of the study area boreal conifers such as Norway spruce (*Picea abies*) and Scots pine (*Pinus sylvestris*) remain dominant but are interspersed with temperate deciduous species like birch (*Betula spp.*) and alder (*Alnus spp.*). These mixed forests represent a gradual transition, where the colder, conifer-reliant ecosystems of the north give way to forests with a higher proportion of temperate broadleaf species in the south. In the region's southernmost nemoral zones, broadleaf deciduous species such as oak (*Quercus spp.*) and beech (*Fagus sylvatica*) become more prominent and often form contiguous forests. This combination results in the study area having a rich mosaic of forest types that blends northern boreal characteristics with elements of temperate deciduous forests.

## Data

### *Swedish National Forest Inventory*

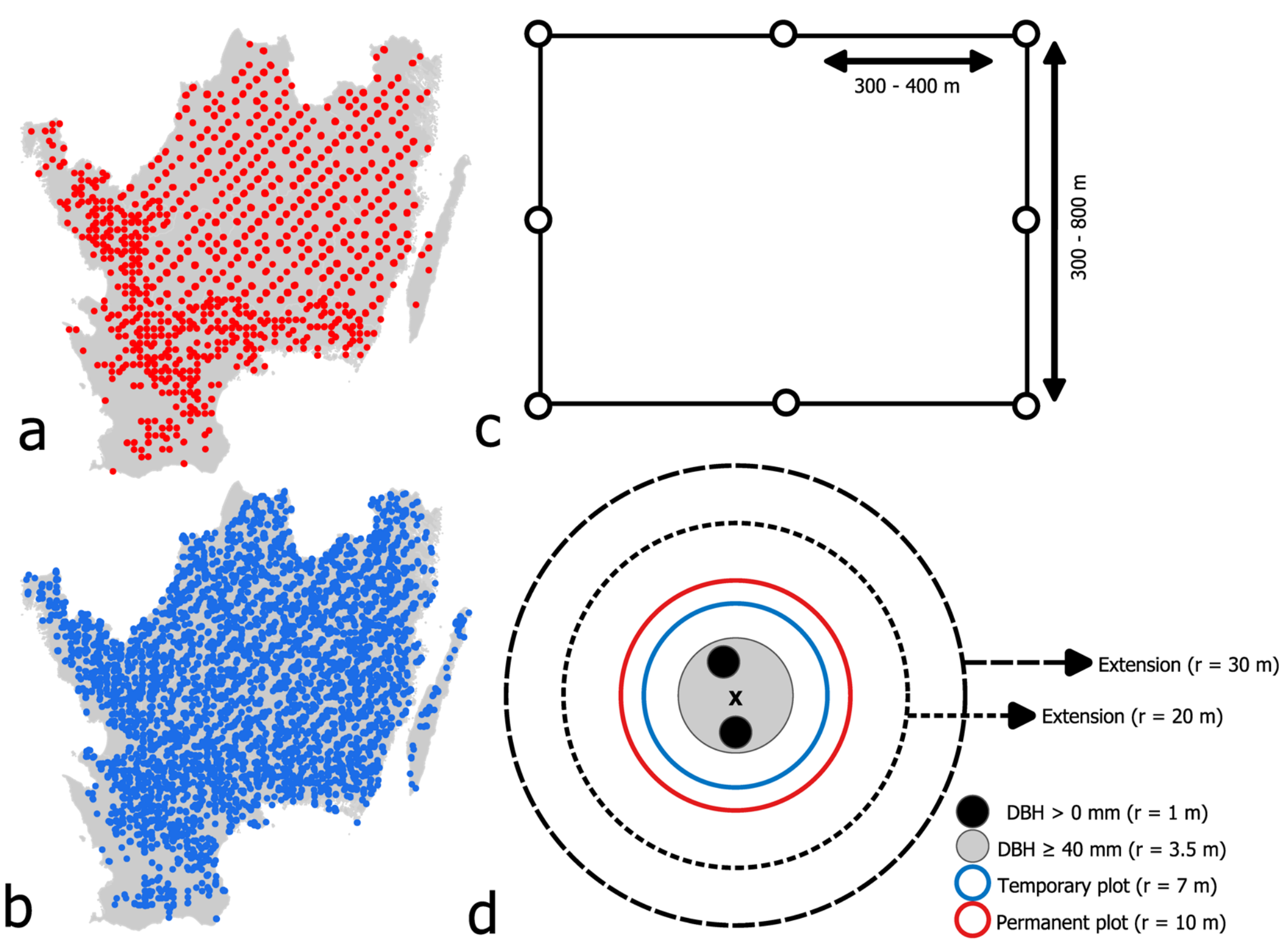

**Figure 2.** Spatial distribution of the permanent **(a)** and temporary tracts **(b)** in the study area that were selected for modelling tree species. Note that the sampling density is greater in the southernmost parts of the study area compared to the northern parts. Each tract takes a square or rectangular shape **(c)** that comprises 4 to 12 plots that are spaced at varying distances. Trees with a diameter of ≥100 mm are measured at breast height (1.3 m), while those with diameters of 40–100 mm and <40 mm are measured in subplots with radii of 3.5 m and 1 m, respectively, within both permanent and temporary plots **(d)**. The extensions around the plots (r = 20 m and r = 30 m) were created through manual photointerpretation to increase the sample size of the Sentinel-1 and Sentinel-2 pixels used to train the ML model.

The Swedish NFI employs probability sampling through a stratified cluster design. Groups of 4 to 12 sample plots, arranged in square or rectangular patterns called tracts, are systematically distributed across the country (Fridman and Westerlund 2016; Fridman et al. 2014). Sixty percent of these plots are permanent and revisited on a five-year rotation, while the remaining 40% are temporary and measured only once to enhance spatial coverage. Permanent plots have an area of 314.16 m² (radius =10 m) (Figure 2 a), while temporary plots have an area 153.94 m² (radius = 7 m) (Figure 2 b). The side lengths of tracts vary depending on location (Figure 2 c). Within the study area, the sides of the tracts are approximately 300 m in the south, increasing to 800 m in the north (reaching 1800 m in the far north of Sweden). Similarly, the spacing between individual plots within tracts varies geographically, starting at 300 m in the south and increasing to 600 m in the north (extending to 800 in the far north of Sweden). Currently, the total sample includes 4,500 permanent tracts and 2,500 temporary tracts (Lundström and Wikberg 2017). Approximately one-fifth of these tracts are surveyed each year, resulting in the measurement of around 12,000 plots annually. All trees that have a diameter of 100 mm or more within each 7 or 10 m radius plot are measured at breast height (DBH, 1.3 m above ground). Trees with DBH between 40 and 100 mm, and those less than 40 mm are measured in subplots (with radii of 3.5 m and 1 m, respectively) within permanent and temporary plots. A subset of sample trees is then chosen with a probability proportional to their basal area at breast height. For these selected sample trees, species and tree class are recorded, and their total height is measured.

## Earth observation data

### Sentinel-1 Synthetic Aperture Radar (SAR)

The Sentinel-1 Level-1 Ground Range Detected (GRD) product was used in this study. This dataset was processed to backscatter coefficients (σ°) and converted to decibels (dB). It also underwent radiometric calibration, orthorectification, and speckle filtering, and was then prepared as an analysis-ready product following Mullissa et al. (2021). The interferometric wide swath acquisition mode was used with dual polarization: VV (vertical transmit, vertical receive) and VH (vertical transmit, horizontal receive). A total of 1,317 Sentinel-1 acquisitions between 2021 and 2023 were processed in GEE. Details of Sentinel-1 GRD are provided in Supplementary Table S1.

### Sentinel-2 Multispectral Imager (MSI)

The Harmonised Sentinel-2 Surface Reflectance (L2A) product used in this study was orthorectified and geometrically corrected using Sen2Cor (Louis et al. 2016; Louis and L2A Team 2021). Sentinel-2 data are provided in three spatial resolutions (10 m, 20 m, 60 m) and thirteen spectral bands. Ten of these bands, blue (B2), green (B3), red (B4), red-edge (B5, B6, B7), near-infrared (B8, B8A), and shortwave-infrared bands (B11, B12), were used in this study. Scenes with less than 70% cloud cover were selected, which resulted in 2,746 Sentinel-2 images from 2021-2023 that were processed in GEE. Details of Sentinel-2 L2A are provided in Supplementary Table S1.

## Auxiliary variables

Geomorphometry influences the spatial distribution of tree species by shaping moisture and nutrient patterns across a landscape. Its effects can be quantified using digital elevation models (DEMs), which provide derivatives such as slope and the topographic wetness index (TWI). Seibert, Stendahl, and Sørensen (2007) linked these variables to variations in organic layer depth, soil acidity, and nutrient gradients. For this study, the FABDEM v1-2 dataset (Forest and Buildings removed Copernicus DEM) (Hawker et al. 2022) was used to derive elevation, slope, aspect, and TWI. FABDEM (Supplementary Figure S2) has been shown to outperform most other broad-scale elevation products, with particularly good performance in forested regions (Osama, Shao, and Freeshah 2023; Huang and Yang 2025; Marsh, Harder, and Pomeroy 2023). FABDEM was resampled and reprojected from 1 arc-second (EPSG:4326) to 10 m (EPSG:3006) using nearest neighbour interpolation to match the spatial resolution and projection of the other datasets in this study.

Canopy height provides information on which tree species are dominant in a forest, since height variation mirror species-specific growth pattern and competitive strategies (Besic et al. 2025). Taller species shape the upper layers of the canopy, and this vertical structuring has been linked to overall species diversity because trees partition light and space differently across heights (Hakkenberg et al. 2016; Farhadur Rahman, Onoda, and Kitajima 2022). Well-known examples include Norway spruce, which can strongly influence the surrounding canopy profile. For this study, canopy height was taken from the map produced by Lang et al. (2023), which was generated using a deep convolutional neural network combined with relative height information derived from the GEDI spaceborne lidar (Dubayah et al. 2020). Its 10 m spatial resolution provides detailed spatial representation of the landscape that closely airborne laser scanning (Torresani et al. 2023). This relatively higher resolution captures height variability more effectively, a feature that has been associated with tree species diversity (Torresani et al. 2020) and improved classification performance (Popova 2025).

# Methods

Data analysis was performed using QGIS 3.34.11, R 4.5 (*terra 1.8, ggpubr 0.6.2, tidyverse 2.0*), Python 3.12 (*scikit-learn 1.6, Earth Engine Python API 1.1.1*) in a Windows 11 Pro environment running on an AMD Ryzen Threadripper PRO workstation with 16 cores, 256 GB RAM and an RTX 4080 GPU.

## *Data preparation*

### *NFI data*

The NFI dataset used in this study had 23,138 plots encompassing the period 1996 – 2023. The dataset contained duplicates from different sampling years and no-data values where species proportions could not be determined. To prepare the data, no-data values were removed, and duplicates were resolved by retaining only the most recent inventory record. Plots that were disturbed by clearcuts (either inside or within a 10-m buffer) were removed by performing spatial and temporal comparisons using the latest version of the Swedish Forest Agency's clearcut database. Plots that were sampled after the most recently recorded clearcut with a mean tree height exceeding 5 meters were kept in order to meet the definition of forest land under the Swedish Forestry Act (Riksdag 1979). To ensure clear species representation and reduce uncertainty in plot locations, only single-species stands were included. A threshold of 80% was applied to differentiate between pure- and mixed-stands (Blickensdörfer et al. 2024; Welle et al. 2022). Therefore, if the basal area of any tree species constituted 80% or more of the composition in a sampling plot, it was classified as a pure-stand plot with that species as the dominant one. The Swedish National Land Cover Database (Naturvårdsverket 2024) was used to remove plots that were not in a productive forest. This procedure yielded a final dataset of 6,527 NFI plots that was used to extract sample pixels for further analysis (Table 1).

Preliminary analysis indicated that the quantity of sample pixels from the final NFI plot dataset was insufficient to produce robust results for some of the deciduous species. To increase the sample size for these rarer species, plots where they were dominant were spatially expanded to areas of 1,256.64 m² (r = 20 m) or 2,827.43 m² (r = 30 m) to include additional Sentinel-1 and Sentinel-2 pixels for model training (Figure 2d). We selected the radius by evaluating expansion feasibility through manual photointerpretation of 1,568 plots dominated by other species using multitemporal high-resolution Google Earth orthophotos (Supplementary Figures S3–S4) and time-stamped Google Street View for plots that were near public roads. If expansion to 30 m would have introduced neighbouring stands or other species into the plot, the radius was limited to 20 m; otherwise, it was extended to 30 m. During photointerpretation, expansion decisions were based on a consistent set of visual criteria to ensure that the expanded area remained representative of the dominant species in the original NFI plot. These criteria included: (1) canopy texture and structural homogeneity (e.g., similar crown size, spacing, and layering across the expanded area), (2) continuity in canopy colour and tonal patterns across seasons in multitemporal imagery, (3) absence of visible stand boundaries or sharp transitions indicative of species changes, and (4) consistent phenological signals (e.g., timing of leaf-out or senescence).

This approach allowed us to enlarge the effective sample size for minority species while maintaining species purity within each expanded plot. It is important to note that the expansion procedure was not applied uniformly across all species. The proportion of plots expanded to 20 m or 30 m varied depending on local stand composition and the feasibility of maintaining single-species dominance during photointerpretation. Consequently, species such as beech and birch, which often occurred in relatively homogeneous stands, had a higher number of expanded plots, and thus sample pixels, whereas species with more heterogeneous or mixed stands, such as alder or aspen, were expanded less frequently (Table 1).

**Table 1.** Summary of the Swedish NFI data (mean ± standard deviation). The samples column represents the number of 10-m pixels extracted from the plot following the expansion procedure (see Methods/Data preparation/NFI data section). Differences between plot and sample counts reflect that the spatial expansion (r = 20 m or 30 m) was applied selectively, depending on stand homogeneity and species dominance.

| Species | Stand Age (Years) | Diameter at breast height, DBH (mm) | Volume ($m^3$ $ha^{-1}$) | Height (m) | Plots | Samples |
|---|---|---|---|---|---|---|
| Scots pine<br>*Pinus sylvestris* | 65 ± 39 | 234 ± 97 | 136 ± 94 | 14 ± 6 | 2209 | 4081 |
| Norway spruce<br>*Picea abies* | 42 ± 25 | 197 ± 92 | 197 ± 152 | 16 ± 7 | 2750 | 5197 |
| Birch<br>*Betula spp.* | 29 ± 23 | 135 ± 100 | 70 ± 61 | 10 ± 7 | 754 | 5421 |
| Oak<br>*Quercus spp.* | 72 ± 42 | 334 ± 179 | 157 ± 140 | 15 ± 8 | 210 | 2348 |
| Alder<br>*Alnus spp.* | 50 ± 27 | 242 ± 102 | 199 ± 174 | 15 ± 7 | 135 | 1028 |
| Beech<br>*Fagus sylvatica* | 84 ± 35 | 387 ± 166 | 256 ± 197 | 21 ± 7 | 287 | 5199 |
| Aspen<br>*Populus spp.* | 37 ± 33 | 195 ± 158 | 143 ± 110 | 12 ± 10 | 46 | 203 |
| Other species*<br>*Other spp.* | 41 ± 35 | 208 ± 132 | 92 ± 91 | 12 ± 9 | 136 | 661 |

* e.g., Goat willow (*Salix caprea*), Larch (*Larix decidua*), European ash (*Fraxinus excelsior*), etc.

### *Satellite and auxiliary Data*

In addition to the raw Sentinel-1 backscatter and Sentinel-2 reflectance, several vegetation indices were derived from them to enhance the discrimination of tree species by capturing key biophysical and phenological characteristics of forest stands. These include the dual-polarization radar vegetation index (Nasirzadehdizaji et al. 2019) (RVIr, Equation 1), cross-ratio index (Blaes et al. 2006) (CR, Equation 2), enhanced vegetation index (Huete et al. 2002) (EVI, Equation 3), and the multispectral ratio vegetation index (Pearson and Miller 1972) (RVIm, Equation 4). Three-month seasonal composites using the median value were generated for the years 2021 to 2023, corresponding to the four seasons: DJF (December–February), MAM (March–May), JJA (June–August), and SON (September–November). A major benefit of deriving STMs from seasonal composites is that they greatly cut processing time while still reflecting key phenological shifts in the forest while reducing the effect of clouds and snow (Nasiri et al. 2023). This approach resulted in 197 features (16 bands and indices × 12 seasons + 5 static auxiliary layers).

$$RVIr = \frac{4\sigma^{o}VH}{(\sigma^{o}VV + \sigma^{o}VH)} \quad (1)$$

$$CR = \frac{\sigma^{o}VH}{\sigma^{o}VV)} \quad (2)$$

$$EVI = 2.5\frac{NIR - RED}{NIR + 6 \times RED - 7.5 \times BLUE + 1} \quad (3)$$

$$RVIm = \frac{RED}{NIR} \quad (4)$$

where, VH (vertical transmit, horizontal receive) and VV (vertical transmit, vertical receive) are Sentinel-1 polarizations; σ° is the backscatter coefficient; BLUE, RED, and NIR are the blue (band 2, 490 nm), red (band 4, 665 nm), and near-infrared (band 8, 842 nm) bands of Sentinel-2. See Supplementary Table S1 for specifics regarding the Sentinel-1 and -2 data.

## *Machine learning (ML) approach*

The final dataset (NFI+Satellite) yielded 24,138 sample pixels (Table 1), which were divided randomly into optimisation and validation subsets. Iterative model training and testing was done on the optimisation subset, and final evaluation was done on the independent validation subset.

### *Class imbalance*

The class distribution in the final dataset was highly imbalanced. An imbalanced dataset causes multi-class classification to be biased toward the majority classes, leading to poor prediction accuracy for underrepresented classes. This happens because the model prioritises minimizing overall error rather than learning meaningful patterns from all classes equally (He and Garcia 2009; Fassnacht et al. 2016). We tested three strategies to remedy class imbalance in our analysis: (1) a simple down-sampling of the dataset using the number of aspen pixels (class with least samples) as a reference resulted in an F1 score of 0.67, (2) merging aspen with the 'other species' class and then using alder as a baseline to down-sample the dataset yielded a slightly higher F1 Score of 0.76, and (3) assigning class weights inversely proportional to class frequencies (Chen et al. 2025), which produced the best results (F1 = 0.83) and was thus implemented in the workflow.

### *Data leakage*

As ML is increasingly being used in ecological remote sensing applications, there is a concern that data leakage could adversely affect the results (Stock, Gregr, and Chan 2023). This occurs due to the presence of spatial autocorrelation when training and validation data are geographically clustered, causing the model to access information that artificially inflates its accuracy (Ploton et al. 2020). We tested three strategies to limit data leakage in our analysis:

(1) the commonly used 80/20 split whereby 80% of the data is assigned to the optimisation subset and 20% to the validation subset while preserving class ratios through stratified sampling resulted in an F1 score of 0.93. Since our data consists of pixels extracted from circular plots, optimisation and validation pixels can come from the same plot, leading to data leakage and over-overoptimistic predictions.

(2) a group-based split that treats each sampling plot as a distinct group and data from the same plot is never in both optimisation and validation. This approach produced relatively poor results (F1 = 0.68) due to the paucity of pixels that can be used for optimisation and validation for rarer species such as aspen.

(3) spatial block partitioning (Valavi et al. 2019; Roberts et al. 2017) that divides the study areas into spatially contiguous blocks of a specified size. Then, entire blocks, rather than individual pixels, are randomly assigned to either the optimisation or validation subset. This ensures that sample pixels located close to one another are not simultaneously used for both model fitting and validation, thereby producing a more realistic assessment of model generalization (Valavi et al. 2019). We tested block sizes of 60 m, 100 m, 150 m, and 200 m, corresponding approximately to increasing degrees of spatial separation among neighbouring plots. Model performance declined at larger block sizes (F1 = 0.82, 0.69, 0.71, and 0.63, respectively) due to reduced training data availability. A 60 m block size was therefore chosen as it provided the best balance between limiting spatial dependence and retaining sufficient data for model calibration.

Given the outcomes of these experiments, we adopted the spatial block partitioning using the 60 m block configuration for model development.

*Algorithm choice*

We initially compared four ML models: one-dimensional convolutional neural network (1D-CNN), support vector machines (SVM), random forest (RF), and extreme gradient boosting (XGB) (Supplementary Table S2). Because the number and influence of hyperparameters varied across models, all four were subjected to the same training-testing partitioning and cross-validation scheme to ensure comparability. We implemented an exhaustive grid search to systematically explore combinations of key hyperparameters within defined ranges for each model and identify the configuration that maximized predictive performance. This approach was selected because it provides a reproducible and unbiased means of evaluating model performance across the full hyperparameter space, ensuring that each algorithm is tuned to its optimal state before comparison. The evaluation metrics included overall accuracy (OA), mean F1 score, precision, recall, and the F1 score per class. Among the tested algorithms, XGB achieved the highest performance (F1 = 0.83, OA = 85%) and was therefore selected for the final tree species classification.

*Hyperparameter optimisation*

XGB has three categories of configuration parameters: general parameters, booster parameters, and learning task parameters (Chen et al. 2025). In this study, we refer to the subset of these parameters that are subject to tuning as *hyperparameters*. Since we used the tree booster (*gbtree*), we focused exclusively on tuning booster hyperparameters related to model complexity and training behaviour. Model performance during tuning was assessed using 10-fold stratified cross-validation that preserved class proportions in each fold. In each iteration, the model was fitted on approximately 17,373 pixels and evaluated on the remaining 1,930 pixels, ensuring that all observations served once as test data.

We used a two-stage hyperparameter tuning strategy beginning with an initial broad search followed by more targeted optimisation. In the first stage, we adopted a stepwise grid search, evaluating combinations around the default XGB settings. The procedure followed a sequential structure: (1) fixing the learning rate at 0.1 while tuning the number of trees (*n_estimators*); (2) adjusting *max_depth* and *min_child_weight*; (3) modifying *min_split_loss*; (4) optimizing *colsample_bytree* and *subsample*; and (5) tuning the regularization terms *reg_alpha* and *reg_lambda*. After narrowing the parameter space, we used Bayesian optimisation (scikit-optimize/BayesSearchCV (Head et al. 2021)) in the second stage, again applying 10-fold cross-validation to identify the most influential parameter combinations within the reduced ranges. Additional tuning of other parameters sometimes improved training performance but slightly reduced test accuracy, suggesting potential overfitting.

The optimised parameters where: *n_estimators* = 3500; *learning_rate* = 0.118914842747; *colsample_bytree* = 0.5; *min_split_loss* = 1e-09; *max_depth* = 4; *min_child_weigh t*= 1; *reg_alpha* = 0.001; *reg_lambda* = 0.01; *subsample* = 0.9. Although this model showed strong performance, we ultimately used a simpler configuration (*n_estimators* = 3000, *learning_rate* = 0.1) to generate the final maps. This decision was based on slightly better testing accuracy and a conscious effort to reduce model complexity and limit the risk of overfitting.

#### *Variable importance*

Variable importance was assessed using total gain, which quantifies the cumulative improvement in the objective function resulting from all splits involving a particular feature across the ensemble of decision trees. Specifically, it sums up the reduction in multiclass log loss achieved at each split where the feature is used. This metric reflects both how often a feature is used in splits and how effective it is in reducing the overall loss, thus presenting a feature's contribution to the model's predictive performance (Chen and Guestrin 2016).

### *Quantifying classification uncertainty*

XGB produces a probability vector for each prediction in a multiclass classification to indicate how likelihood of each class for a given input. In such probability distributions, entropy (also called Shannon's entropy (Shannon 1948)) provides a theoretically grounded framework for quantifying uncertainty (Wang 2008). In simple terms, entropy is a measure of disorder or randomness in a system. It quantifies how 'spread out' the predictions are. The entropy H of a predicted class probability vector $p = [p_1, p_2, ..., p_K]$, where K is the number of classes, is calculated as (Equation 5):

$$H_{(p)} = -\sum_{i=1}^{K} p_i \log(p_i) \qquad (5)$$

Then, dividing by the maximum possible entropy, *log(K)*, normalises the values and ensures values fall between 0 and 1. Entropy values close to 0 indicate high confidence, because most of the probability mass is assigned to a single class. For example, a probability vector that looks like this: [0.93, 0.02, 0.01, 0.00, 0.01, 0.01, 0.02] has very low entropy because the classifier is clearly favouring the first class (0.93) over the others. In contrast, a vector such as [0.15, 0.14, 0.14, 0.14, 0.14, 0.14, 0.15] distributes probabilities more evenly across all classes, indicating high uncertainty, and consequently high entropy, close to *log(K)*. Thus, a continuous-valued uncertainty map is generated by applying this entropy calculation to every pixel to accompany the discrete multiclass predictions, which helps highlight model confidence.

## Results

### *Classification performance*

The final map (Figure 3 a) based on XGB and the 60 m spatial block partitioning had an overall accuracy of 85% and a Matthews correlation coefficient of 0.81 based on the independent validation subset. On average, the model achieved a mean recall of 0.80±0.09 and mean precision of 0.84±0.05, with a corresponding F1 score of 0.82±0.07 (Table 2). The average omission error was 0.19±0.09, and commission error was 0.15±0.05, reflecting relatively balanced misclassification rates across the classes. The 10-fold cross-validation accuracies, which are based on optimisation subset, were all above the 0.9 mark: 0.93±0.02 (precision), 0.95±0.03 (recall), and 0.94±0.02 (F1 score) (Supplementary Figure S5). The model also demonstrated strong overall performance on the independent validation subset across most tree species classes, with particularly high scores for dominant species. Beech performed best overall, with the highest recall (0.93) and a strong F1 score of 0.92, indicating that the model reliably identified this species with low omission and commission errors. Two of the most spatially widespread species in Götaland, Norway spruce and Scots pine, showed similarly solid performance, with F1 scores of 0.82 and 0.83 respectively. Birch also stood out with near-identical recall and precision values, reflecting near-equal accuracy in identifying this class. Oak showed a high precision (0.88) and recall (0.86), resulting in an F1 score of 0.87. Aspen had slightly lower recall (0.76) but very high precision (0.92), suggesting it was occasionally missed by the model but almost always correctly classified when predicted. In contrast, alder and 'other species' showed the weakest performance as both classes had relatively low recall (0.66 and 0.66, respectively), indicating a higher rate of omission errors. Alder's F1 score was 0.72, and 'other species' was 0.71, which is still reasonably high but notably lower than the other classes and likely due to more heterogeneous or less distinct spectral signatures.

### *Model uncertainty*

Although the normalized entropy-based uncertainty was overall low across the study area (0.16±0.19, Figure 3 b), there were notable differences in classification confidence across tree species (Figure 4). Norway spruce, Scots pine, and beech exhibit the lowest median entropy values, indicating that the model consistently assigned these species with high confidence across the study area. These species likely had distinct and stable spectral-temporal signatures, which contributed to more certain predictions. In contrast, species such as aspen, alder, and other species show considerably higher entropy values, reflected by wider interquartile ranges and higher medians. These higher uncertainty levels suggest greater confusion during classification, likely due to a combination of spectral similarity with other deciduous species, more heterogeneous stand compositions, and lower representation in the training data. The aspen class in particular displays the greatest spread and central tendency in entropy, probably due to the small number of sample pixels in the final dataset (203). The 'other species' class has the second-highest uncertainty despite having more than three times the number of samples as aspen (661), and this high uncertainty is consistent with the challenge of lumping diverse minor species into a single group, which reduces the model's ability to make confident predictions. Intermediate uncertainty is observed for birch and oak, which show moderate median entropy values. Although generally well-classified based on performance metrics, these species may exhibit spectral overlap with others under certain seasonal or structural conditions, which contributes to uncertainty in predictions.

### *Variable importance*

The variable importance plot for the final XGB model (Figure 5) reveals that spectral indices and selected Sentinel-2 bands were the most influential predictors for tree species classification. Among these, the multispectral ratio vegetation index (RVIm) had the highest contribution, followed closely by Sentinel-2 band 11 (B11) and the enhanced vegetation index (EVI). These variables provided significant gains in predictive performance, particularly during spring and summer (Supplementary Figure S6), suggesting that seasonal phenological differences play a key role in distinguishing tree species. Other important optical bands in the top ten included B7, B4, B8, and B12, which also showed strong contributions across multiple seasons, especially spring and summer. The VV polarization from Sentinel-1 data and the radar vegetation index (RVIr) were also among the top ten contributors, indicating the added value of radar-derived variables in complementing optical data, especially in less optimal atmospheric conditions or for detecting canopy structure. Vegetation indices and spectral bands from autumn and winter generally showed lower overall importance but still contributed meaningfully to predictors like EVI, B4, B8, and radar-derived indices such as RVIr and VH, indicating that incorporating year-round observations enhances model robustness. The contribution of winter data, while lower overall, was still visible for EVI (Supplementary Figure S6). This likely reflects its utility in identifying evergreen species such as Norway spruce and Scots pine, which maintain spectral and structural signatures during the dormant season. In contrast to the optical and radar data, auxiliary variables such as canopy height (CHM), slope, elevation (DEM), aspect, and topographic wetness (TWI) had relatively low importance scores. Among them, CHM was the most influential, though still much less so than the leading spectral and backscatter predictors. This suggests that while elevation and terrain-based predictors offer some explanatory power, they play a secondary role compared to spectral and radar information in this study area.

### *Comparison with official statistics*

Figure 6 shows the comparison between tree species area estimates from the final XGB model and official data from Swedish Forest Statistics (Riksskogstaxeringen 2024). Across all the six counties that cover the study area (Blekinge, Halland, Jönköping, Kalmar, Kronoberg, and Skåne) the model's predictions show strong agreement with the reference statistics, as indicated by the high Spearman's rho ($\rho = 0.94$, $P < 0.001$). This suggests that the model scales effectively to broader regional assessments. At the individual county level, most counties exhibited high correlation ($\rho \geq 0.8$) and statistically significant relationships ($P < 0.005$) (Supplementary Figure S7) between the mapped and reported species cover, particularly in Halland, Skåne, and Blekinge, where Spearman's correlation values were the highest.

**Table 2.** Performance metrics of the final XGB map based on the spatially independent validation set created using the block partitioning.

| **Tree Species Class** | **Performance Metrics** | | | | |
|---|---|---|---|---|---|
| | **Recall** *Producer's Accuracy* | **Precision** *User's Accuracy* | **Omission Error** | **Commission Error** | **F1 Score** |
| Norway spruce | 0.846 | 0.806 | 0.154 | 0.193 | 0.826 |
| Scots pine | 0.829 | 0.831 | 0.171 | 0.169 | 0.830 |
| Birch | 0.849 | 0.853 | 0.151 | 0.147 | 0.851 |
| Beech | 0.936 | 0.919 | 0.063 | 0.081 | 0.928 |
| Oak | 0.868 | 0.880 | 0.132 | 0.120 | 0.874 |
| Alder | 0.663 | 0.788 | 0.336 | 0.212 | 0.721 |
| Aspen | 0.764 | 0.928 | 0.235 | 0.071 | 0.839 |
| Other species | 0.661 | 0.769 | 0.338 | 0.231 | 0.711 |
| **Mean ± StdDev** | **0.802 ± 0.09** | **0.847 ± 0.05** | **0.197 ± 0.09** | **0.153 ± 0.05** | **0.823 ± 0.07** |
| **Overall accuracy = 85%** <br> **Matthews correlation coefficient = 0.81** | | | | | |

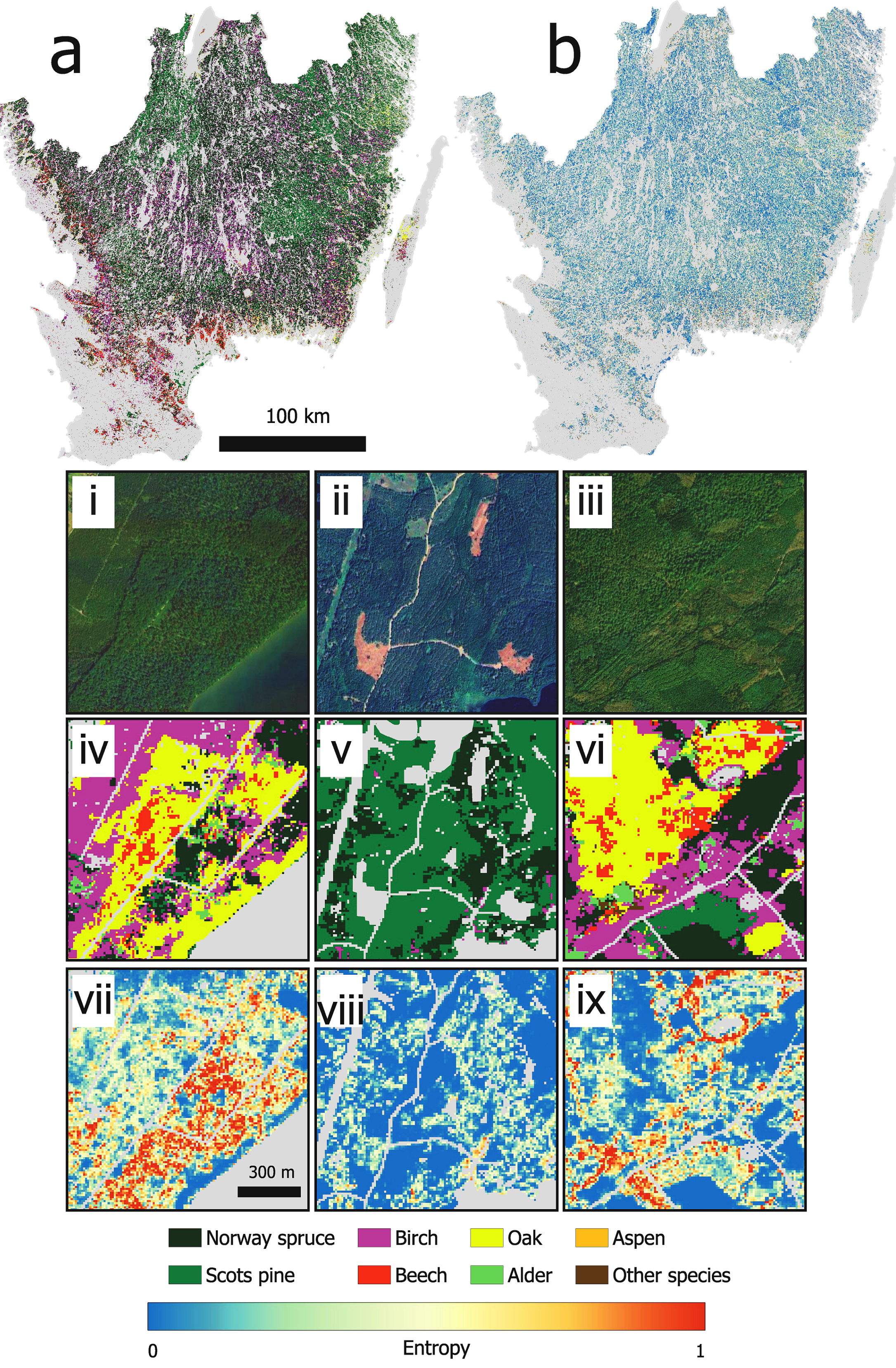

**Figure 3.** Final tree species classification (a) and the entropy-based uncertainty map (b). The six stacked panels show map details for the classification and uncertainty for: part of a ~200-year-old oak forest at Visingsö in Jönköping (i, iv, vii); a typical Scots pine and Norway spruce dominated production forest in Kronoberg (ii, v, vii); a nemoral forest in Skåne made up of a mosaic of stands comprising different tree species (iii, vi, ix). Normalized entropy-based uncertainty values range from 0 (low uncertainty / high confidence, one class

dominates) to 1 (high uncertainty / low confidence, probabilities evenly spread across classes).

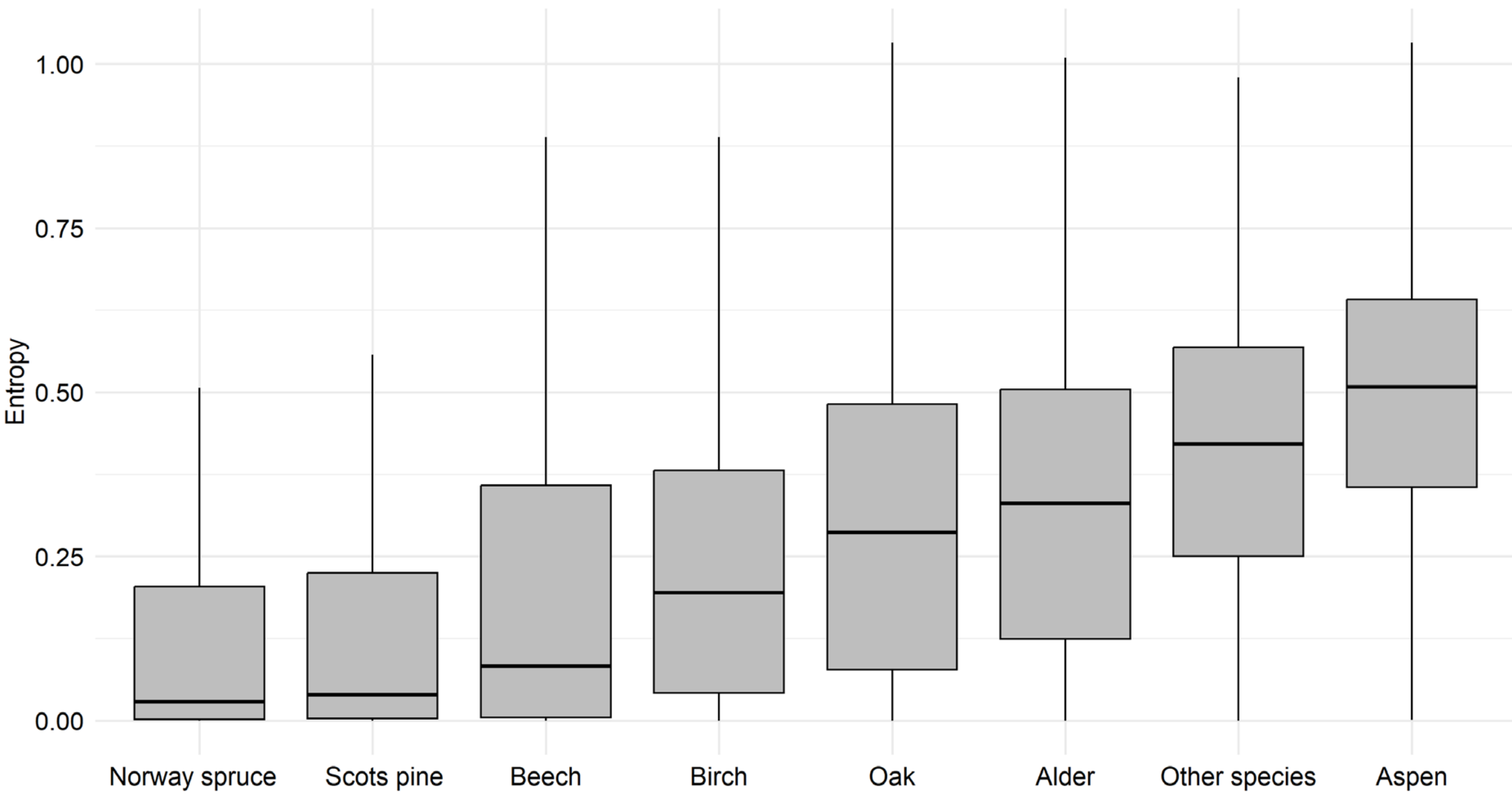

**Figure 4.** Boxplots showing the distribution of the normalized entropy across tree species classes in the final XGB model. Lower entropy values indicate higher model confidence in assigning a given class. Species with higher variability and median entropy values, such as aspen and 'other species', reflect greater prediction uncertainty, while species like Norway spruce and Scots pine exhibit confident classifications.

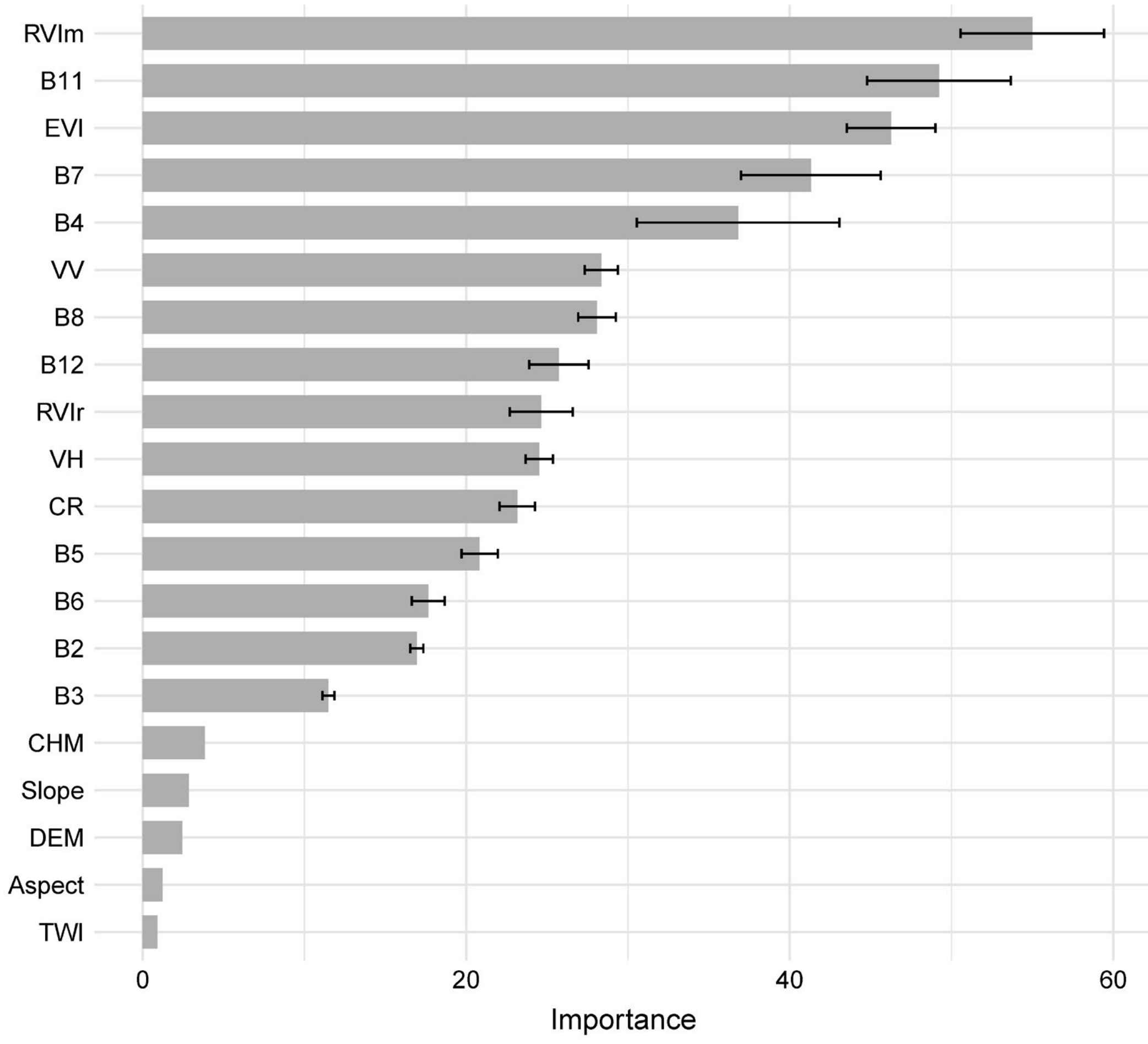

**Figure 5.** Variable importance scores from the XGB model, aggregated using total gain. This metric reflects how much each predictor variable contributes to improving the model's predictions by reducing the difference between predicted and actual tree species. Higher values indicate more influential variables. Error bars represent the variability of each variable's importance across seasons. See Supplementary Figure S6 for a seasonal disaggregation of variable importance.

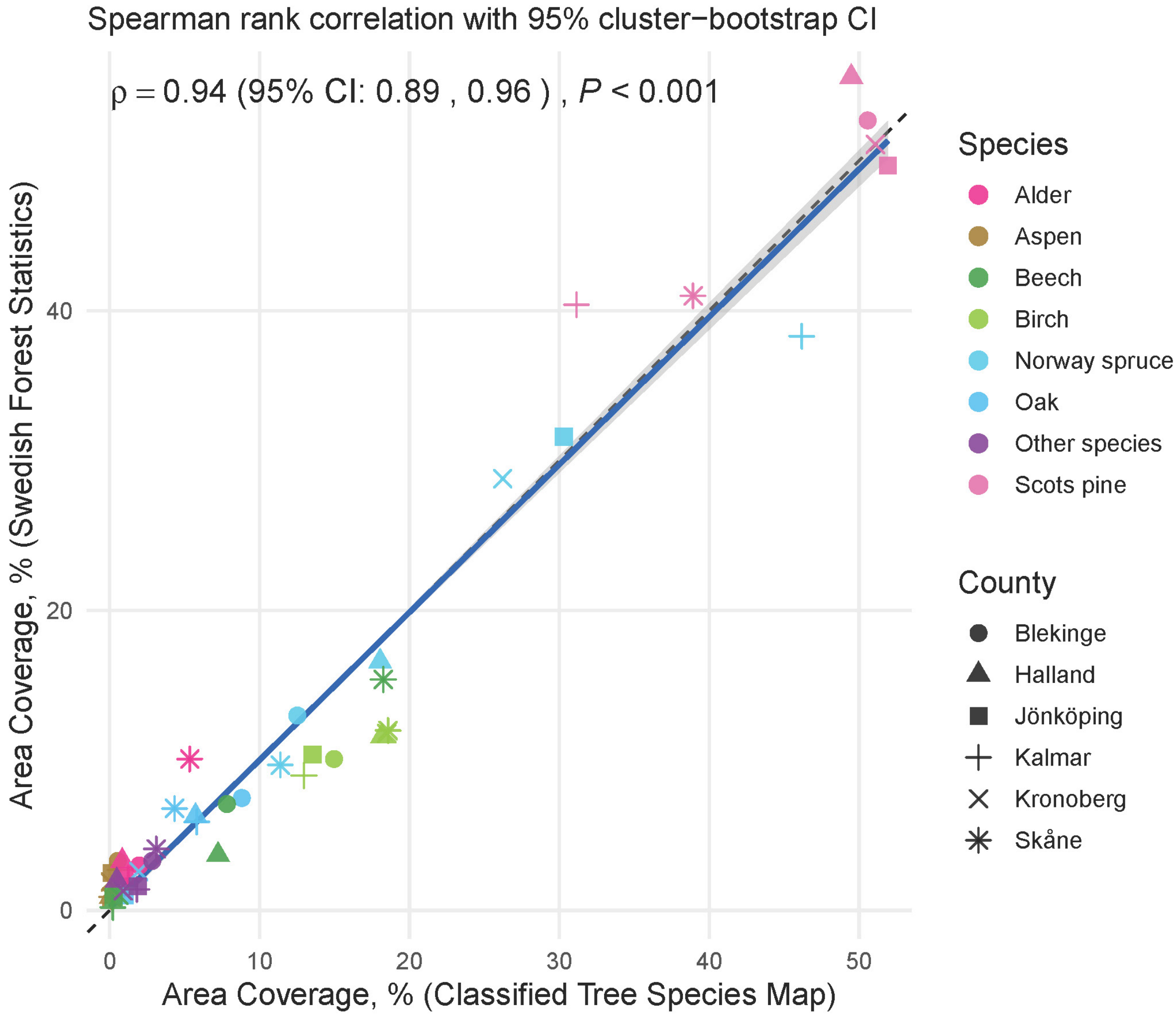

**Figure 6.** Comparison of the mapped percent cover of the tree species with the official Swedish forest statistics (Riksskogstaxeringen 2024). See Supplementary Figure S7 for a disaggregation of this figure into each county.

## Discussion

Mapping tree species across large areas has become increasingly feasible due to advancements in remote sensing technologies and is an active area of research (Pu 2021; Immitzer and Atzberger 2023). This study demonstrates how the complementary use of Sentinel-1 radar, Sentinel-2 multispectral data, and auxiliary variables can support detailed, spatially continuous mapping of dominant tree species across southern Sweden, with associated pixel-level uncertainty. The final product should be interpreted as a map of the most likely dominant species per 10 m pixel rather than a full representation of mixed-species composition. We achieved an overall accuracy of 85% (F1 = 0.82) and strong agreement with official forest statistics (ρ = 0.94) using an XGB classifier trained on NFI plots. These results are broadly consistent with previous work in European forests (Blickensdörfer et al. 2024; Grabska-Szwagrzyk et al. 2024; Breidenbach et al. 2021) that demonstrated the strength of satellite time series for distinguishing dominant species. However, while this approach reduces label uncertainty and enhances separability, it also means that the reported accuracy primarily reflects model performance in well-defined, monospecific stands. This is consistent with the findings from D'Amico et al. (2024) showing that dominant-species training often leads to reduced generalizability in heterogeneous forests. Mixed or transitional canopies, common in many parts of the study area, are inherently harder to classify at 10-m resolution, and model performance in such areas is expected to be lower than the aggregate accuracy implies. These considerations highlight the importance of the uncertainty layer, which provides additional guidance when interpreting predictions in heterogeneous forests. This study therefore extends earlier work by not only mapping dominant species but also quantifying the spatial variation in model confidence.

### *Model uncertainty*

The entropy-based uncertainty map (Figure 3 b) revealed clear species-specific differences in model confidence that closely mirrored the classification performance metrics. Species with high F1 scores such as beech (F1 = 0.92), Norway spruce (F1 = 0.82), and Scots pine (F1 = 0.83) also exhibited low median entropy values (Figure 4). These species likely benefit from distinct spectral-temporal signatures, relatively small intra-species variability of the remote sensing signal, and sufficient representation in the training data. In the case of beech, its low uncertainty likely reflects both its phenological distinctiveness and its structural characteristics in the study area. Beech stands in southern Sweden are generally older, taller, and have higher basal area and volume than the other broadleaved species included in this study (Table 1). These structural traits tend to produce denser, more cohesive canopies with pronounced seasonal dynamics, especially during leaf-out and senescence, which are well captured by red-edge and SWIR-sensitive features (Morley et al. 2020). As a result, beech exhibits a more stable spectral-temporal profile than species such as birch or oak, both of which show greater within-class variability in stand age, height, and volume (Table 1). Birch, in particular, is characterised by younger and more heterogeneous stands, while oak displays comparatively large variation in DBH and stand structure. These factors contribute to more variable canopy reflectance and align with the moderate entropy values observed for these species.

In contrast, species such as alder (F1 = 0.72) and the aggregated 'other species' class (F1 = 0.71) had the highest uncertainty values, reflected in both wider interquartile ranges and higher medians in the entropy distribution. This pattern aligns with their relatively lower precision and recall scores, suggesting that the model struggled to reliably distinguish these classes, likely due to greater spectral overlap, more heterogeneous stand composition, or limited training samples. Lu et al. (2022) used entropy to quantify uncertainty in land-cover maps produced through upscaling, demonstrating that it captures categorical variation well and pinpoints areas of high uncertainty, especially near class edges. Higher entropy values were associated with more ambiguous or underrepresented classes, which supports our interpretation that it provides complementary information to accuracy metrics. This is further supported by Peters et al. (2009) who used entropy to assess prediction uncertainty in vegetation distribution models. Their findings revealed a strong link between entropy and model uncertainty, where higher entropy marked ambiguous predictions, often arising from limited input information or tightly mixed species groups. This reinforces our finding that entropy is not only a reflection of model performance but also a valuable diagnostic for evaluating confidence in predictions, helping to identify areas and species where additional scrutiny is needed.

### *Class performance*

The strong performance across dominant tree species such as beech, Norway spruce, Scots pine, and birch suggests that the model successfully captured key phenological and structural signatures using seasonal composites. Species with distinct seasonal dynamics, such as beech (a deciduous tree with strong spring and autumn spectral transitions), showed very high classification certainty and F1 scores above 0.9. Conversely, uncertainty and classification errors were higher for rarer species like alder and the aggregated 'other species' class, consistent with prior findings that rare or mixed-species stands introduce confusion (Fassnacht et al. 2016; Grabska-Szwagrzyk et al. 2024). Figure 4 further supports this, showing clear gradients in entropy across species classes, reinforcing the idea that class rarity and within-class spectral variability are key determinants of model uncertainty.

The challenge of within-class spectral variability is well-documented in large-scale forest classification, where differences in canopy structure, site conditions, and background signals often lead to increased intra-species variability of the remote sensing signals and classification errors (Fassnacht et al. 2016). We trained a single model using a diverse set of features, including vegetation indices, radar cross-polarizations, and seasonal composites, which resulted in higher model confidence for well-represented species. This supports the idea that even with a diverse set of features, sufficient training data are required for ML algorithms to capture consistent spectral behaviour and structural traits to enhance the classification reliability of tree species. Furthermore, the manual expansion of plots for rare species may also introduce a subtle selection bias, as expansions were only applied when the surrounding canopy appeared visually homogeneous and clearly dominated by the target species. This approach ensured label purity but may preferentially sample well-defined, uniform stands, potentially leading to slightly optimistic accuracy estimates for species that otherwise occur in more heterogeneous conditions. Although the overall influence on model performance is likely limited, given the small number of expanded plots relative to the full dataset, it remains an important consideration. Future work could mitigate this bias by incorporating automated homogeneity metrics, probabilistic labelling approaches, or larger reference datasets that reduce the need for manual expansion.

Classes with lower representation in the training data, such as aspen (n = 203) and the heterogeneous ‘other species’ group (n = 661), were more prone to misclassification and exhibited higher uncertainty. This is in line with earlier studies that highlight the sensitivity of classification accuracy to reference data imbalance (He and Garcia 2009), particularly for rare or spectrally ambiguous species (Fassnacht et al. 2016). We attempted to address this issue by applying class weighting within the XGB model, which improved the average F1 score compared to other techniques such as down-sampling. This was done by manipulating the *sample_weight* parameter in XGB to assign different weights to individual training samples, influencing how much each sample contributes to the model's loss function and gradient computation during training. The approach should be useful in scenarios like handling class imbalance by giving higher weights to minority class samples to prevent the model from being biased toward the majority class (Chen et al. 2025; Liu and Zhou 2013; Ferrario and Hämmerli 2019). However, the entropy-based uncertainty revealed that these minority classes continued to pose challenges. For example, aspen had high precision (0.92), moderate recall (0.76), and a relatively high median entropy (0.51±0.20) despite addressing class imbalance. This suggests that while the model was selective and usually correct when predicting aspen, it remained uncertain in many of these predictions. The high entropy suggests feature overlap with similar deciduous species (e.g., birch or alder), making it difficult for the model to confidently separate aspen despite weighting adjustments. In contrast, alder and the “other species” class showed both lower recall (0.66 for both) and moderate precision (0.78 and 0.76, respectively), reflecting more heterogeneous stand structures and greater within-class spectral variability.

### *Variable importance*

The variable importance analysis (Figure 5) highlights the critical role of optical predictors in distinguishing tree species across the study area, with results largely consistent with other remote sensing studies in European forests (Hemmerling, Pflugmacher, and Hostert 2021; Grabska et al. 2019). Among the Sentinel-2 variables, shortwave infrared (SWIR) band 11 and vegetation indices such as the multispectral ratio vegetation index (RVIm) and enhanced vegetation index (EVI) emerged as particularly influential. The high importance of SWIR (Band 11) can be attributed to its sensitivity to canopy moisture content and leaf internal structure (Lukeš et al. 2013), both of which vary across species due to differences in water retention strategies, leaf morphology, and wood anatomy (Ustin and Gamon 2010). Similarly, EVI and RVIm, both responsive to canopy greenness, have been shown to reflect phenological and structural traits that help distinguish broadleaf from coniferous species (Peng et al. 2017; Gao et al. 2023). These indices are particularly responsive during leaf-on periods in spring and summer, which is reflected in their seasonal contribution patterns (Supplementary Figure S6). Radar-based variables derived from Sentinel-1 also played an important role, with VV backscatter and the radar vegetation index (RVIr) ranking among the top ten predictors. VV backscatter responds to surface roughness and the vertical elements of the canopy, which makes it especially helpful for recognising structurally complex conifers like Norway spruce and Scots pine (Bjerreskov, Nord-Larsen, and Fensholt 2021; Dostálová et al. 2021). RVIr combines both VV and VH polarizations allowing it to capture variation in volume scattering and canopy density, which are important features in mixed stands or where crown layers overlap (Schellenberg et al. 2023; Hu et al. 2024).

Although optical and radar predictors dominated the variable importance rankings, geomorphometric variables, elevation, slope, aspect, and TWI, contributed little to model performance. Given the relatively low topographic variation in the study area, the gradients that could aid species discrimination are likely weak (Waser et al. 2021; Yu et al. 2020; Zhang et al. 2024). The resampling of FABDEM to 10 m may further reduce the already limited terrain signal, which together likely explains the low importance of topographic predictors relative to the Sentinel-1/2 features. While the overall impact on classification accuracy appears minimal, future work could assess whether higher-resolution DEMs or lidar-derived terrain products enhance species discrimination in more topographically complex landscapes. CHM showed moderate influence, which is likely due to its ability to differentiate the growth forms of tree species (Torresani et al. 2023; Zhang et al. 2025), such as old beech stands or young birch plantations.

### *Seasonal dynamics*

Seasonally aggregated variable importance scores showed statistically significant differences across seasons (Kruskal–Wallis, $P = 0.0016$; Supplementary Fig. S8). Imagery acquired during periods of active canopy development, spring (MAM) and summer (JJA), provided the highest predictive value. These seasonal patterns align with earlier findings (Axelsson et al. 2021; Persson, Lindberg, and Reese 2018) that reported that spring and summer Sentinel-2 observations provide the most discriminative information for separating tree species in Swedish forests. This reflects the strong phenological and structural contrasts among species during leaf-out and peak growth phases. In contrast, autumn (SON) offered only moderate importance, likely because senescence drives several deciduous species toward more similar spectral states, reducing inter-species separability (Kozhoridze et al. 2016). Winter (DJF) contributed the least, which is expected given widespread leaf-off conditions, reduced pigment activity, and muted phenological differences among broadleaved species. Evergreen conifers retain some discriminative signal during winter, but this advantage is likely not sufficient to offset the overall reduction in canopy-level variability across the landscape (Wang, Springer, and Gamon 2023). These patterns show that most species-specific information is concentrated in spring and summer (Lechner et al. 2022). Autumn and winter play a secondary role, and their inclusion helps stabilise the model by capturing year-round conditions even though they add limited extra power for distinguishing species.

### *Comparison with official forest statistics*

The comparison between mapped species coverage and official forest statistics from the Swedish government (Figure 6 and Supplementary Figure S7) shows strong overall agreement, but some regional discrepancies are evident. In particular, ‘other species’ and alder tend to be slightly overestimated in counties such as Kalmar and Skåne, while species like birch are marginally underestimated in Kronoberg and Halland. These deviations, although modest in magnitude, likely reflect several interacting sources of classification uncertainty. First, spectral similarity among broadleaf species (Fassnacht et al. 2016), such as alder, birch, and aspen, can lead to confusion in the classifier as these species often share similar phenological profiles and canopy structure, particularly when imaged at 10-meter resolution. Second, official statistics are often based on plot-level NFI data that may aggregate or apportion species differently than pixelwise satellite classification (Barrett et al. 2016). For example, mixed stands with co-dominant broadleaf species may be recorded under a single dominant species in the NFI, while the classifier distributes class labels on a per-pixel basis, potentially resulting in discrepancies in aggregate cover estimates. Third, limitations in spatial resolution and canopy penetration can lead to underrepresentation of species present in the sub-canopy or in structurally complex stands (Dalponte and Coomes 2016). This is especially relevant for species like birch that may occur as a secondary canopy layer beneath taller spruce or pine, which dominate the spectral signal in much of the study area. Despite these challenges, the high degree of alignment across counties confirms the map’s overall reliability and suggests that most deviations can be attributed to predictable sources of ecological and remote sensing complexity rather than model failure.

The strong correlation ($\rho = 0.94$) between the mapped species cover and official forest statistics indicates that aggregated area estimates are less sensitive to localised classification errors. Whereas pixel-level metrics penalise each misclassification equally, even when confusion is among spectrally similar or spatially adjacent species, area-based statistics smooth out many of these inconsistencies. For example, if some pixels of birch are misclassified as alder, but alder pixels are also occasionally misclassified as birch, the aggregate area estimates for both species may still remain close to their true values at the county level. This effect is particularly relevant in southern Sweden, where species like birch, alder, and aspen often co-occur and may be spectrally similar, especially at 10-m resolution. Similarly, Breidenbach et al. (2021) demonstrated that Sentinel-2-based maps of dominant tree species in Norway showed strong agreement with NFI data. They showed that aggregated species area estimates at the national level largely resulted in improvements compared to field-based estimates alone.

Furthermore, differences in how species dominance is interpreted, per pixel in remote sensing versus per stand in field-based inventories, can lead to small discrepancies that likely do not have significant impacts on total area estimates. As noted by Wulder et al. (2020) in their Landsat-based assessment of Canadian forest area, such structural and definitional mismatches between field inventories and raster-based classifications can still yield high agreement when results are aggregated spatially. Their study reported 84% spatial agreement between Landsat-derived and NFI-derived forest area classifications, and only 3% difference in total forest area at the national level, albeit local classification discrepancies were non-negligible.

### *Advancing operational forest monitoring*

The integration of EO data, NFI field observations, and ML offers a powerful framework for advancing operational forest monitoring. Satellite time series provide the spatial continuity and temporal frequency needed to track forest composition and structural change, while NFI plots supply the high-quality, standardised field measurements required for model calibration and validation. ML acts as the link between these datasets, enabling the extraction of species-specific signals from large volumes of EO data and supporting automated, repeatable workflows that can be scaled across regions and updated as new data become available. Combining these complementary components allows the production of consistent, spatially explicit indicators, such as dominant species maps and their associated uncertainty layers that can support routine assessments of forest condition, harvesting impacts, regeneration patterns, and management outcomes.

This combined approach also aligns with documented trends in NFIs globally, where most countries now rely on EO data to enhance estimation and produce spatial products such as forest type or species-group maps (Barrett et al. 2016). Many NFIs report that remotely sensed information has become essential for improving the precision of area estimates, supporting small-area assessments, and reducing the cost and logistical burdens associated with field-only approaches (Barrett et al. 2016). The inclusion of an entropy-based uncertainty map further enhances operational utility. High-uncertainty areas can be prioritised for field verification, targeted NFI re-measurements, or additional remote sensing analysis, while low-uncertainty areas can be used with greater confidence in applications such as forest planning, habitat assessments, or summary reporting. Uncertainty can also be propagated through downstream analyses by down-weighting uncertain pixels in regional estimates of species composition, biomass, or carbon stocks. As satellite archives expand and cloud-based computation makes large-scale processing more accessible, ML-enabled EO-NFI integration provides a practical foundation for operational forest monitoring systems capable of delivering timely, scalable information for decision-making.

### *Study limitations*

The temporal mismatch between NFI plot measurements (1996-2023) and satellite composites (2021-2023) remains a source of uncertainty. Although plots affected by recent clearcuts were removed, stands can still change structurally between measurement and image acquisition, potentially altering canopy height, density, and phenology while species identity remains unchanged (Zhirin, Knyazeva, and Eydlina 2014; Wu et al. 2018; Croft, Chen, and Noland 2014). Such discrepancies introduce variation into the training data, which may reduce class separability for younger or actively managed stands. However, a small amount of such noise may not necessarily degrade model performance. In some cases, limited label noise can act as a regulariser and improve generalization by preventing models from overfitting to highly specific conditions (Frenay and Verleysen 2014; Zhang et al. 2021). In this study, the mismatch could introduce either a small regularizing effect or a reduction in species separability, but its net influence cannot be isolated because plot ages and structural conditions evolve independently of image acquisition. Addressing this will require future studies designed explicitly to test how temporal offsets between field and satellite data affect model behaviour.

The feature set excluded explicitly spatial predictors such as the coordinates, distance to nearest features, etc. Although we included 197 spectral, radar, and topographic variables, additional spatial descriptors could have captured residual spatial structure not fully represented by multi-sensor metrics and potentially help explain the reduced performance under spatial block cross-validation. Incorporating ecologically meaningful spatial features may therefore improve generalisation (Hermosilla, Wulder, et al. 2022). Training relied on pure stands ($\geq$ 80% dominance), which lowers label uncertainty but limits exposure to the variability present in mixed-species stands (Fassnacht et al. 2016). As a result, predictions in heterogeneous or transitional stands should be interpreted cautiously. Residual spatial autocorrelation also remains a concern. Despite using spatial block splitting, neighbouring pixels may still share similar environmental or management conditions, potentially leading to mildly optimistic performance estimates. Exploring larger block sizes or fully independent spatial holdouts could help better quantify model robustness.

## Conclusions

Accurate, spatially explicit information on forest composition is increasingly vital for managing biodiversity, monitoring ecological change, and informing sustainable forest policy. This study presents a scalable approach for mapping dominant tree species in Sweden using freely available Sentinel-1 and Sentinel-2 data and Swedish National Forest Inventory field observations in a machine learning framework. The classification model, based on extreme gradient boosting with Bayesian optimisation, achieved high accuracy, with F1 scores exceeding 0.8 for most species. Our mapped species distributions closely matched official forest statistics, with a correlation coefficient of 0.94, despite modest pixel-level uncertainty. Much of this uncertainty was concentrated in classes with limited training data or greater spectral overlap, such as aspen and the other species class. Entropy-based uncertainty was used to quantify classification confidence at the pixel level, which provided a practical tool for assessing spatial variation in model reliability. Our model identified key predictors capturing species-specific differences in seasonal spectral response and canopy structure. Features derived from optical vegetation indices and bands contributed most to classification accuracy, while radar-derived metrics provided complementary information on forest structure. The inclusion of these features enabled the model to capture both coniferous and broadleaf species patterns across the ecologically heterogeneous landscapes of southern Sweden. As the need for timely and consistent forest information grows, the workflow developed here provides a foundation for producing species-level maps with explicit uncertainty estimates. The approach is well suited for operational monitoring contexts and can support applications ranging from forest management planning to ecological assessments.

## References

Atzberger, C , G  Zeug, P  Defourny, L  Aragão, L  Hammarström, and M Immitzer. 2020. "Monitoring of forests through remote sensing " In. Directorate-General for Environment: European Commission.

Axelsson, Arvid, Eva Lindberg, Heather Reese, and Håkan Olsson. 2021. "Tree species classification using Sentinel-2 imagery and Bayesian inference." *International Journal of Applied Earth Observation and Geoinformation* 100:102318. https://doi.org/https://doi.org/10.1016/j.jag.2021.102318.

Barrett, Frank, Ronald E. McRoberts, Erkki Tomppo, Emil Cienciala, and Lars T. Waser. 2016. "A questionnaire-based review of the operational use of remotely sensed data by national forest inventories." *Remote Sensing of Environment* 174:279-289. https://doi.org/https://doi.org/10.1016/j.rse.2015.08.029.

Besic, N., N. Picard, C. Vega, J. D. Bontemps, L. Hertzog, J. P. Renaud, F. Fogel, et al. 2025. "Remote-sensing-based forest canopy height mapping: some models are useful, but might they provide us with even more insights when combined?" *Geosci. Model Dev.* 18 (2):337-359. https://doi.org/10.5194/gmd-18-337-2025.

Bjerreskov, Kristian Skau, Thomas Nord-Larsen, and Rasmus Fensholt. 2021. "Classification of Nemoral Forests with Fusion of Multi-Temporal Sentinel-1 and 2 Data." *Remote Sensing* 13 (5):950.

Blaes, X., P. Defourny, U. Wegmuller, A. Della Vecchia, L. Guerriero, and P. Ferrazzoli. 2006. "C-band polarimetric indexes for maize monitoring based on a validated radiative transfer model." *IEEE Transactions on Geoscience and Remote Sensing* 44 (4):791-800. https://doi.org/10.1109/TGRS.2005.860969.

Blickensdörfer, Lukas, Katja Oehmichen, Dirk Pflugmacher, Birgit Kleinschmit, and Patrick Hostert. 2024. "National tree species mapping using Sentinel-1/2 time series and German National Forest Inventory data." *Remote Sensing of Environment* 304:114069. https://doi.org/https://doi.org/10.1016/j.rse.2024.114069.

Breidenbach, Johannes, Lars T. Waser, Misganu Debella-Gilo, Johannes Schumacher, Johannes Rahlf, Marius Hauglin, Stefano Puliti, and Rasmus Astrup. 2021. "National mapping and estimation of forest area by dominant tree species using Sentinel-2 data." *Canadian Journal of Forest Research* 51 (3):365-379. https://doi.org/10.1139/cjfr-2020-0170.

Brockerhoff, Eckehard G., Luc Barbaro, Bastien Castagneyrol, David I. Forrester, Barry Gardiner, José Ramón González-Olabarria, Phil O'B Lyver, et al. 2017. "Forest biodiversity, ecosystem functioning and the provision of ecosystem services." *Biodiversity and Conservation* 26 (13):3005-3035. https://doi.org/10.1007/s10531-017-1453-2.

Cesaretti, Lorenzo, Saverio Francini, Davide Travaglini, and Piermaria Corona. 2025. "Copernicus for monitoring European forests: strengths, challenges and potential." *European Journal of Remote Sensing* 58 (1):2540109. https://doi.org/10.1080/22797254.2025.2540109.

Chen, T, T He, M Benesty, V Khotilovich, Y Tang, H Cho, K Chen, et al. 2025. "xgboost: Extreme Gradient Boosting." In. https://xgboost.readthedocs.io/en/stable/: R package version 3.0.0.1

Chen, Tianqi, and Carlos Guestrin. 2016. "XGBoost: A Scalable Tree Boosting System." In *Proceedings of the 22nd ACM SIGKDD International Conference on Knowledge Discovery and Data Mining*, 785–794. San Francisco, California, USA: Association for Computing Machinery.

Creed, Irena F., Marian Weber, Francesco Accatino, and David P. Kreutzweiser. 2016. "Managing Forests for Water in the Anthropocene—The Best Kept Secret Services of Forest Ecosystems." *Forests* 7 (3):60.

Croft, H., J. M. Chen, and T. L. Noland. 2014. "Stand age effects on Boreal forest physiology using a long time-series of satellite data." *Forest Ecology and Management* 328:202-208. https://doi.org/https://doi.org/10.1016/j.foreco.2014.05.023.

D'Amico, Giovanni, Mats Nilsson, Arvid Axelsson, and Gherardo Chirici. 2024. "Data homogeneity impact in tree species classification based on Sentinel-2 multitemporal data case study in central Sweden." *International Journal of Remote Sensing* 45 (15):5050-5075. https://doi.org/10.1080/01431161.2024.2371082.

Dalponte, Michele, and David A. Coomes. 2016. "Tree-centric mapping of forest carbon density from airborne laser scanning and hyperspectral data." *Methods in Ecology and Evolution* 7 (10):1236-1245. https://doi.org/https://doi.org/10.1111/2041-210X.12575.

Dlamini, Cliff S. 2019. "Contribution of Forest Ecosystem Services Toward Food Security and Nutrition." In *Zero Hunger*, edited by Walter Leal Filho, Anabela Marisa Azul, Luciana Brandli, Pinar Gökcin Özuyar and Tony Wall, 1-18. Cham: Springer.

Dostálová, Alena, Mait Lang, Janis Ivanovs, Lars T. Waser, and Wolfgang Wagner. 2021. "European Wide Forest Classification Based on Sentinel-1 Data." *Remote Sensing* 13 (3):337.

Dubayah, Ralph, James Bryan Blair, Scott Goetz, Lola Fatoyinbo, Matthew Hansen, Sean Healey, Michelle Hofton, et al. 2020. "The Global Ecosystem Dynamics Investigation: High-resolution laser ranging of the Earth's forests and topography." *Science of Remote Sensing* 1:100002. https://doi.org/https://doi.org/10.1016/j.srs.2020.100002.

European Commission. 2021. "New EU Forest Strategy for 2030." In.

Farhadur Rahman, Md, Yusuke Onoda, and Kaoru Kitajima. 2022. "Forest canopy height variation in relation to topography and forest types in central Japan with LiDAR." *Forest Ecology and Management* 503:119792. https://doi.org/https://doi.org/10.1016/j.foreco.2021.119792.

Fassnacht, Fabian Ewald, Hooman Latifi, Krzysztof Stereńczak, Aneta Modzelewska, Michael Lefsky, Lars T. Waser, Christoph Straub, and Aniruddha Ghosh. 2016. "Review of studies on tree species classification from remotely sensed data." *Remote Sensing of Environment* 186:64-87. https://doi.org/https://doi.org/10.1016/j.rse.2016.08.013.

Ferrario, Andrea, and Roger Hämmerli. 2019. "On Boosting: Theory and Applications." In.: ETH Zurich, Mobiliar Lab for Analytics.

Frenay, B., and M. Verleysen. 2014. "Classification in the Presence of Label Noise: A Survey." *IEEE Transactions on Neural Networks and Learning Systems* 25 (5):845-869. https://doi.org/10.1109/TNNLS.2013.2292894.

Fridman, Jonas, Sören Holm, Mats Nilsson, Per Nilsson, Anna Hedström Ringvall, and Göran Ståhl. 2014. *Adapting National Forest Inventories to changing requirements – the case of the Swedish National Forest Inventory at the turn of the 20th century*. Vol. 48.

Fridman, Jonas, and Bertil Westerlund. 2016. "Swedish National Forest Inventory." In *National Forest Inventories: Assessment of Wood Availability and Use*, edited by Claude Vidal, Iciar A. Alberdi, Laura Hernández Mateo and John J. Redmond, 769-782. Cham: Springer International Publishing.

Gao, Si, Run Zhong, Kai Yan, Xuanlong Ma, Xinkun Chen, Jiabin Pu, Sicong Gao, Jianbo Qi, Gaofei Yin, and Ranga B. Myneni. 2023. "Evaluating the saturation effect of vegetation indices in forests using 3D radiative transfer simulations and satellite observations." *Remote Sensing of Environment* 295:113665. https://doi.org/https://doi.org/10.1016/j.rse.2023.113665.

Gorelick, Noel, Matt Hancher, Mike Dixon, Simon Ilyushchenko, David Thau, and Rebecca Moore. 2017. "Google Earth Engine: Planetary-scale geospatial analysis for everyone." *Remote Sensing of Environment* 202:18-27. https://doi.org/https://doi.org/10.1016/j.rse.2017.06.031.

Grabska-Szwagrzyk, E., D. Tiede, M. Sudmanns, and J. Kozak. 2024. "Map of forest tree species for Poland based on Sentinel-2 data." *Earth Syst. Sci. Data* 16 (6):2877-2891. https://doi.org/10.5194/essd-16-2877-2024.

Grabska, Ewa, Patrick Hostert, Dirk Pflugmacher, and Katarzyna Ostapowicz. 2019. "Forest Stand Species Mapping Using the Sentinel-2 Time Series." *Remote Sensing* 11 (10):1197.

Hakkenberg, Christopher R., Conghe Song, Robert K. Peet, and Peter S. White. 2016. "Forest structure as a predictor of tree species diversity in the North Carolina Piedmont." *Journal of Vegetation Science* 27 (6):1151-1163. https://doi.org/https://doi.org/10.1111/jvs.12451.

Hawker, Laurence, Peter Uhe, Luntadila Paulo, Jeison Sosa, James Savage, Christopher Sampson, and Jeffrey Neal. 2022. "A 30 m global map of elevation with forests and buildings removed." *Environmental Research Letters* 17 (2):024016. https://doi.org/10.1088/1748-9326/ac4d4f.

He, H., and E. A. Garcia. 2009. "Learning from Imbalanced Data." *IEEE Transactions on Knowledge and Data Engineering* 21 (9):1263-1284. https://doi.org/10.1109/TKDE.2008.239.

Head, Tim, Manoj Kumar, Holger Nahrstaedt, Gilles Louppe, and Iaroslav Shcherbatyi. 2021. "scikit-optimize/scikit-optimize (v0.9.0)." *Zenodo*.

Hemmerling, Jan, Dirk Pflugmacher, and Patrick Hostert. 2021. "Mapping temperate forest tree species using dense Sentinel-2 time series." *Remote Sensing of Environment* 267:112743. https://doi.org/https://doi.org/10.1016/j.rse.2021.112743.

Hermosilla, Txomin, Alex Bastyr, Nicholas C. Coops, Joanne C. White, and Michael A. Wulder. 2022. "Mapping the presence and distribution of tree species in Canada's forested ecosystems." *Remote Sensing of Environment* 282:113276. https://doi.org/https://doi.org/10.1016/j.rse.2022.113276.

Hermosilla, Txomin, Michael A. Wulder, Joanne C. White, and Nicholas C. Coops. 2022. "Land cover classification in an era of big and open data: Optimizing localized implementation and training data selection to improve mapping outcomes." *Remote Sensing of Environment* 268:112780. https://doi.org/https://doi.org/10.1016/j.rse.2021.112780.

Hu, Xueqian, Li Li, Jianxi Huang, Yelu Zeng, Shuo Zhang, Yiran Su, Yujiao Hong, and Zixiang Hong. 2024. "Radar vegetation indices for monitoring surface vegetation: Developments, challenges, and trends." *Science of The Total Environment* 945:173974. https://doi.org/https://doi.org/10.1016/j.scitotenv.2024.173974.

Huang, Jiapeng, and Xiaozhu Yang. 2025. "Evaluation and improvement of the vertical accuracy of the global open DEM under forest environment." *Geocarto International* 40 (1):2453024. https://doi.org/10.1080/10106049.2025.2453024.

Huete, A., K. Didan, T. Miura, E. P. Rodriguez, X. Gao, and L. G. Ferreira. 2002. "Overview of the radiometric and biophysical performance of the MODIS vegetation indices." *Remote Sensing of Environment* 83 (1):195-213. https://doi.org/https://doi.org/10.1016/S0034-4257(02)00096-2.

Immitzer, Markus, and Clement Atzberger. 2023. "Tree Species Diversity Mapping—Success Stories and Possible Ways Forward." *Remote Sensing* 15 (12):3074.

Joppa, Lucas. 2019. "A planetary computer to avert environmental disaster." *Scientific American* 19.

Kozhoridze, Giorgi, Nikolai Orlovsky, Leah Orlovsky, Dan G. Blumberg, and Avi Golan-Goldhirsh. 2016. "Remote sensing models of structure-related biochemicals and pigments for classification of trees." *Remote Sensing of Environment* 186:184-195. https://doi.org/https://doi.org/10.1016/j.rse.2016.08.024.

Lang, Nico, Walter Jetz, Konrad Schindler, and Jan Dirk Wegner. 2023. "A high-resolution canopy height model of the Earth." *Nature Ecology & Evolution* 7 (11):1778-1789. https://doi.org/10.1038/s41559-023-02206-6.

Lechner, Michael, Alena Dostálová, Markus Hollaus, Clement Atzberger, and Markus Immitzer. 2022. "Combination of Sentinel-1 and Sentinel-2 Data for Tree Species Classification in a Central European Biosphere Reserve." *Remote Sensing* 14 (11):2687.

Liang, Jingjing, Thomas W. Crowther, Nicolas Picard, Susan Wiser, Mo Zhou, Giorgio Alberti, Ernst-Detlef Schulze, et al. 2016. "Positive biodiversity-productivity relationship predominant in global forests." *Science* 354 (6309):aaf8957. https://doi.org/doi:10.1126/science.aaf8957.

Liu, Xiaojuan, Stefan Trogisch, Jin-Sheng He, Pascal A. Niklaus, Helge Bruelheide, Zhiyao Tang, Alexandra Erfmeier, et al. 2018. "Tree species richness increases ecosystem carbon storage in subtropical forests." *Proceedings of the Royal Society B: Biological Sciences* 285 (1885):20181240. https://doi.org/doi:10.1098/rspb.2018.1240.

Liu, Xu-Ying, and Zhi-Hua Zhou. 2013. "Ensemble Methods for Class Imbalance Learning." In *Imbalanced Learning*, 61-82.

Louis, Jérôme, Vincent Debaecker, Bringfried Pflug, Magdalena Main-Knorn, Jakub Bieniarz, Uwe Mueller-Wilm, Enrico Cadau, and Ferran Gascon. 2016. Sentinel-2 Sen2Cor: L2A processor for users. Paper presented at the Proceedings of the Living Planet Symposium 2016, Prague.

Louis, Jérôme, and L2A Team. 2021. "Sentinel-2 Level-2A Algorithm Theoretical Basis Document." In.: European Space Agency.

Lu, Yunduo, Sun Peijun, Linghu Linna, and Meng and Zhang. 2022. "Uncertainty evaluation approach based on Shannon entropy for upscaled land use/cover maps." *Journal of Land Use Science* 17 (1):648-657. https://doi.org/10.1080/1747423X.2022.2141364.

Lukeš, Petr, Stenberg Pauline, Rautiainen Miina, Mõttus Matti, and Kalle M. and Vanhatalo. 2013. "Optical properties of leaves and needles for boreal tree species in Europe." *Remote Sensing Letters* 4 (7):667-676. https://doi.org/10.1080/2150704X.2013.782112.

Lundström, Anders, and Per-Erik Wikberg. 2017. "Sweden." In *Forest Inventory-based Projection Systems for Wood and Biomass Availability*, edited by Susana Barreiro, Mart-Jan Schelhaas, Ronald E. McRoberts and Gerald Kändler, 289-301. Cham: Springer International Publishing.

Marsh, Christopher B., Phillip Harder, and John W. Pomeroy. 2023. "Validation of FABDEM, a global bare-earth elevation model, against UAV-lidar derived elevation in a complex forested mountain catchment." *Environmental Research Communications* 5 (3):031009. https://doi.org/10.1088/2515-7620/acc56d.

Morin, Xavier, Lorenz Fahse, Hervé Jactel, Michael Scherer-Lorenzen, Raúl García-Valdés, and Harald Bugmann. 2018. "Long-term response of forest productivity to climate change is mostly driven by change in tree species composition." *Scientific Reports* 8 (1):5627. https://doi.org/10.1038/s41598-018-23763-y.

Morin, Xavier, Lorenz Fahse, Michael Scherer-Lorenzen, and Harald Bugmann. 2011. "Tree species richness promotes productivity in temperate forests through strong complementarity between species." *Ecology Letters* 14 (12):1211-1219. https://doi.org/https://doi.org/10.1111/j.1461-0248.2011.01691.x.

Morley, Peter J., Alistair S. Jump, Martin D. West, and Daniel N. M. Donoghue. 2020. "Spectral response of chlorophyll content during leaf senescence in European beech trees." *Environmental Research Communications* 2 (7):071002. https://doi.org/10.1088/2515-7620/aba7a0.

Mullissa, Adugna, Andreas Vollrath, Christelle Odongo-Braun, Bart Slagter, Johannes Balling, Yaqing Gou, Noel Gorelick, and Johannes Reiche. 2021. "Sentinel-1 SAR Backscatter Analysis Ready Data Preparation in Google Earth Engine." *Remote Sensing* 13 (10). https://doi.org/10.3390/rs13101954.

Nasiri, Vahid, Mirela Beloiu, Ali Asghar Darvishsefat, Verena C. Griess, Carmen Maftei, and Lars T. Waser. 2023. "Mapping tree species composition in a Caspian temperate mixed forest based on spectral-temporal metrics and machine learning." *International Journal of Applied Earth Observation and Geoinformation* 116:103154. https://doi.org/https://doi.org/10.1016/j.jag.2022.103154.

Nasirzadehdizaji, Rouhollah, Fusun Balik Sanli, Saygin Abdikan, Ziyadin Cakir, Aliihsan Sekertekin, and Mustafa Ustuner. 2019. "Sensitivity Analysis of Multi-Temporal Sentinel-1 SAR Parameters to Crop Height and Canopy Coverage." *Applied Sciences* 9 (4):655.

Naturvårdsverket. "National Land Cover Database (Nationella Marktäckedata, NMD)." https://www.naturvardsverket.se/verktyg-och-tjanster/kartor-och-karttjanster/nationella-marktackedata/.

Nitoslawski, S. A., K. Wong-Stevens, J. W. N. Steenberg, K. Witherspoon, L. Nesbitt, and C. C. Konijnendijk van den Bosch. 2021. "The Digital Forest: Mapping a Decade of Knowledge on Technological Applications for Forest Ecosystems." *Earth's Future* 9 (8):e2021EF002123. https://doi.org/https://doi.org/10.1029/2021EF002123.

Osama, Nahed, Zhenfeng Shao, and Mohamed Freeshah. 2023. "The FABDEM Outperforms the Global DEMs in Representing Bare Terrain Heights." *Photogrammetric Engineering & Remote Sensing* 89 (10):613-624.

Ottosen, Thor-Bjørn, Geoffrey Petch, Mary Hanson, and Carsten A. Skjøth. 2020. "Tree cover mapping based on Sentinel-2 images demonstrate high thematic accuracy in Europe." *International Journal of Applied Earth Observation and Geoinformation* 84:101947. https://doi.org/https://doi.org/10.1016/j.jag.2019.101947.
Pan, Yude, Richard A. Birdsey, Oliver L. Phillips, Richard A. Houghton, Jingyun Fang, Pekka E. Kauppi, Heather Keith, et al. 2024. "The enduring world forest carbon sink." *Nature* 631 (8021):563-569. https://doi.org/10.1038/s41586-024-07602-x.
Patacca, Marco, Marcus Lindner, Manuel Esteban Lucas-Borja, Thomas Cordonnier, Gal Fidej, Barry Gardiner, Ylva Hauf, et al. 2023. "Significant increase in natural disturbance impacts on European forests since 1950." *Global Change Biology* 29 (5):1359-1376. https://doi.org/https://doi.org/10.1111/gcb.16531.
Pearson, Robert Lawrence, and Lee Durward Miller. 1972. Remote mapping of standing crop biomass for estimation of the productivity of the shortgrass prairie, Pawnee National Grasslands, Colorado. Paper presented at the Proceedings of the 8th International Symposium on Remote Sensing of the Environment
Peh, Kelvin S.-H. , Richard T. Corlett, and Yves Bergeron. 2025. "Part 3: Forest Flora and Fauna." In *Routledge Handbook of Forest Ecology*, 211 - 339. London: Routledge.
Peng, Dailiang, Chaoyang Wu, Cunjun Li, Xiaoyang Zhang, Zhengjia Liu, Huichun Ye, Shezhou Luo, Xinjie Liu, Yong Hu, and Bin Fang. 2017. "Spring green-up phenology products derived from MODIS NDVI and EVI: Intercomparison, interpretation and validation using National Phenology Network and AmeriFlux observations." *Ecological Indicators* 77:323-336. https://doi.org/https://doi.org/10.1016/j.ecolind.2017.02.024.
Persson, Magnus, Eva Lindberg, and Heather Reese. 2018. "Tree Species Classification with Multi-Temporal Sentinel-2 Data." *Remote Sensing* 10 (11):1794.
Peters, Jan, Niko E. C. Verhoest, Roeland Samson, Marc Van Meirvenne, Liesbet Cockx, and Bernard De Baets. 2009. "Uncertainty propagation in vegetation distribution models based on ensemble classifiers." *Ecological Modelling* 220 (6):791-804. https://doi.org/https://doi.org/10.1016/j.ecolmodel.2008.12.022.
Ploton, Pierre, Frédéric Mortier, Maxime Réjou-Méchain, Nicolas Barbier, Nicolas Picard, Vivien Rossi, Carsten Dormann, et al. 2020. "Spatial validation reveals poor predictive performance of large-scale ecological mapping models." *Nature Communications* 11 (1):4540. https://doi.org/10.1038/s41467-020-18321-y.
Popova, Anastasia. 2025. "Improving the Accuracy of Tree Species Mapping by Sentinel-2 Images Using Auxiliary Data—A Case Study of Slyudyanskoye Forestry Area near Lake Baikal." *Forests* 16 (3):487.
Preislerová, Zdenka, Corrado Marcenò, Javier Loidi, Gianmaria Bonari, Dariia Borovyk, Rosario G. Gavilán, Valentin Golub, et al. 2024. "Structural, ecological and biogeographical attributes of European vegetation alliances." *Applied Vegetation Science* 27 (1):e12766. https://doi.org/https://doi.org/10.1111/avsc.12766.
Pu, Ruiliang. 2021. "Mapping Tree Species Using Advanced Remote Sensing Technologies: A State-of-the-Art Review and Perspective." *Journal of Remote Sensing* 2021. https://doi.org/doi:10.34133/2021/9812624.
Puletti, N, F Chianucci, and C Castaldi. 2018. "Use of Sentinel-2 for forest classification in Mediterranean environments." *Annals of Silvicultural Research* 42 (1):32–38.
Raši, Rastislav. 2020. "State of Europe's Forests 2020." In.: Ministerial Conference on the Protection of Forests in Europe.
Riksdag. 1979. "Skogsvårdslag." In *1979:429*, 2 § första stycket 2. Sverige: Landsbygds- och infrastrukturdepartementet.
Riksskogstaxeringen. 2024. "Forest Statistics 2024 - Official Statistics of Sweden." In. https://www.slu.se/globalassets/ew/org/centrb/rt/dokument/skogsdata/skogsdata_2024_web.pdf.
Roberge, J-M., C. Fries, E. Normark, E. Mårald, A. Sténs, C. Sandström, J. Sonesson, C. Appelqvist, and T. Lundmark. 2020. "Forest management in Sweden - Current practice and historical background." In.: Skogsstyrelsen.

Roberts, David R., Volker Bahn, Simone Ciuti, Mark S. Boyce, Jane Elith, Gurutzeta Guillera-Arroita, Severin Hauenstein, et al. 2017. "Cross-validation strategies for data with temporal, spatial, hierarchical, or phylogenetic structure." *Ecography* 40 (8):913-929. https://doi.org/https://doi.org/10.1111/ecog.02881.

Schadauer, Klemens, Rasmus Astrup, Johannes Breidenbach, Jonas Fridman, Stephan Gräber, Michael Köhl, Kari T. Korhonen, et al. 2024. "Access to exact National Forest Inventory plot locations must be carefully evaluated." *New Phytologist* 242 (2):347-350. https://doi.org/https://doi.org/10.1111/nph.19564.

Schellenberg, Konstantin, Thomas Jagdhuber, Markus Zehner, Sören Hese, Marcel Urban, Mikhail Urbazaev, Henrik Hartmann, Christiane Schmullius, and Clémence Dubois. 2023. "Potential of Sentinel-1 SAR to Assess Damage in Drought-Affected Temperate Deciduous Broadleaf Forests." *Remote Sensing* 15 (4):1004.

Seibert, Jan, Johan Stendahl, and Rasmus Sørensen. 2007. "Topographical influences on soil properties in boreal forests." *Geoderma* 141 (1):139-148. https://doi.org/https://doi.org/10.1016/j.geoderma.2007.05.013.

Senf, Cornelius, and Rupert Seidl. 2021. "Mapping the forest disturbance regimes of Europe." *Nature Sustainability* 4 (1):63-70. https://doi.org/10.1038/s41893-020-00609-y.

Shannon, C. E. 1948. "A Mathematical Theory of Communication." *Bell System Technical Journal* 27 (3):379-423. https://doi.org/https://doi.org/10.1002/j.1538-7305.1948.tb01338.x.

Stock, Andy, Edward J. Gregr, and Kai M. A. Chan. 2023. "Data leakage jeopardizes ecological applications of machine learning." *Nature Ecology & Evolution* 7 (11):1743-1745. https://doi.org/10.1038/s41559-023-02162-1.

Torresani, Michele, Duccio Rocchini, Alessandro Alberti, Vítězslav Moudrý, Michael Heym, Elisa Thouverai, Patrick Kacic, and Enrico Tomelleri. 2023. "LiDAR GEDI derived tree canopy height heterogeneity reveals patterns of biodiversity in forest ecosystems." *Ecological Informatics* 76:102082. https://doi.org/https://doi.org/10.1016/j.ecoinf.2023.102082.

Torresani, Michele, Duccio Rocchini, Ruth Sonnenschein, Marc Zebisch, Heidi C. Hauffe, Michael Heym, Hans Pretzsch, and Giustino Tonon. 2020. "Height variation hypothesis: A new approach for estimating forest species diversity with CHM LiDAR data." *Ecological Indicators* 117:106520. https://doi.org/https://doi.org/10.1016/j.ecolind.2020.106520.

Udali, Alberto, Emanuele Lingua, and Henrik J. Persson. 2021. "Assessing Forest Type and Tree Species Classification Using Sentinel-1 C-Band SAR Data in Southern Sweden." *Remote Sensing* 13 (16):3237.

Ustin, Susan L., and John A. Gamon. 2010. "Remote sensing of plant functional types." *New Phytologist* 186 (4):795-816. https://doi.org/https://doi.org/10.1111/j.1469-8137.2010.03284.x.

Valavi, Roozbeh, Jane Elith, José J. Lahoz-Monfort, and Gurutzeta Guillera-Arroita. 2019. "blockCV: An r package for generating spatially or environmentally separated folds for k-fold cross-validation of species distribution models." *Methods in Ecology and Evolution* 10 (2):225-232. https://doi.org/https://doi.org/10.1111/2041-210X.13107.

Viana-Soto, A., and C. Senf. 2025. "The European Forest Disturbance Atlas: a forest disturbance monitoring system using the Landsat archive." *Earth Syst. Sci. Data* 17 (6):2373-2404. https://doi.org/10.5194/essd-17-2373-2025.

Vorovencii, Iosif, Lucian Dincă, Vlad Crişan, Ruxandra-Georgiana Postolache, Codrin-Leonid Codrean, Cristian Cătălin, Constantin Irinel Greşiţă, Sanda Chima, and Ion Gavrilescu. 2023. "Local-scale mapping of tree species in a lower mountain area using Sentinel-1 and -2 multitemporal images, vegetation indices, and topographic information." *Frontiers in Forests and Global Change* Volume 6 - 2023. https://doi.org/10.3389/ffgc.2023.1220253.

Wang, Qiuping A. 2008. "Probability distribution and entropy as a measure of uncertainty." *Journal of Physics A: Mathematical and Theoretical* 41 (6):065004. https://doi.org/10.1088/1751-8113/41/6/065004.

Wang, Ran, Kyle R. Springer, and John A. Gamon. 2023. "Confounding effects of snow cover on remotely sensed vegetation indices of evergreen and deciduous trees: An experimental study." *Global Change Biology* 29 (21):6120-6138. https://doi.org/https://doi.org/10.1111/gcb.16916.

Waser, Lars T., Marius Rüetschi, Achilleas Psomas, David Small, and Nataliia Rehush. 2021. "Mapping dominant leaf type based on combined Sentinel-1/-2 data – Challenges for mountainous countries." *ISPRS Journal of Photogrammetry and Remote Sensing* 180:209-226. https://doi.org/https://doi.org/10.1016/j.isprsjprs.2021.08.017.

Wegler, Marco, Patrick Kacic, Frank Thonfeld, Stefanie Holzwarth, Niklas Jaggy, Ursula Gessner, and Claudia Kuenzer. 2025. "Tree species from space: a new product for Germany based on Sentinel-1 and -2 time series." *International Journal of Remote Sensing* 46 (16):6075-6108. https://doi.org/10.1080/01431161.2025.2530236.

Welle, Torsten, Lukas Aschenbrenner, Kevin Kuonath, Stefan Kirmaier, and Jonas Franke. 2022. "Mapping Dominant Tree Species of German Forests." *Remote Sensing* 14 (14):3330.

Wu, Qiaoli, Conghe Song, Jinling Song, Jindi Wang, Shaoyuan Chen, and Bo Yu. 2018. "Impacts of Leaf Age on Canopy Spectral Signature Variation in Evergreen Chinese Fir Forests." *Remote Sensing* 10 (2):262. https://doi.org/10.3390/rs10020262.

Wulder, Michael A, Txomin Hermosilla, Graham Stinson, François A Gougeon, Joanne C White, David A Hill, and Byron P Smiley. 2020. "Satellite-based time series land cover and change information to map forest area consistent with national and international reporting requirements." *Forestry: An International Journal of Forest Research* 93 (3):331-343. https://doi.org/10.1093/forestry/cpaa006.

Yu, Xiaozhi, Dengsheng Lu, Xiandie Jiang, Guiying Li, Yaoliang Chen, Dengqiu Li, and Erxue Chen. 2020. "Examining the Roles of Spectral, Spatial, and Topographic Features in Improving Land-Cover and Forest Classifications in a Subtropical Region." *Remote Sensing* 12 (18):2907.

Zhang, Jia, Hao Li, Jia Wang, Yuying Liang, Rui Li, and Xiaoting Sun. 2024. "Exploring the Differences in Tree Species Classification between Typical Forest Regions in Northern and Southern China." *Forests* 15 (6):929.

Zhang, Lei, Liu Yang, Jinhua Sun, Qimeng Zhu, Ting Wang, and Hui Zhao. 2025. "Estimation of Tree Species Diversity in Warm Temperate Forests via GEDI and GF-1 Imagery." *Forests* 16 (4):570.

Zhang, Shengmin, Dries Landuyt, Kris Verheyen, and Pieter De Frenne. 2022. "Tree species mixing can amplify microclimate offsets in young forest plantations." *Journal of Applied Ecology* 59 (6):1428-1439. https://doi.org/https://doi.org/10.1111/1365-2664.14158.

Zhang, Yikai, Songzhu Zheng, Pengxiang Wu, Mayank Goswami, and Chao Chen. 2021. "Learning with feature-dependent label noise: A progressive approach." *ArXiv Preprints*.

Zhirin, V. M., S. V. Knyazeva, and S. P. Eydlina. 2014. "Dynamics of spectral brightness of the species/age structure for groups of forest types on Landsat satellite images." *Contemporary Problems of Ecology* 7 (7):788-796. https://doi.org/10.1134/S1995425514070142.